\documentclass[a4paper,twocolumn,11pt,accepted=2024-09-04]{quantumarticle}
\pdfoutput=1
\usepackage[utf8]{inputenc}
\usepackage[english]{babel}
\usepackage[T1]{fontenc}
\usepackage{amsmath}
\usepackage{hyperref}

\usepackage{tikz}
\usepackage{lipsum}

\usepackage{enumerate}
\usepackage{soul}
\usepackage[utf8]{inputenc}
\usepackage[T1]{fontenc}
\usepackage{float}
\usepackage{physics}
\usepackage{amsmath}
\usepackage{nccmath}
\usepackage{mathtools,amsfonts,amssymb,amsthm,bm}
\usepackage{color}
\usepackage{xcolor}

\usepackage{graphicx}
\usepackage{subfigure}
\usepackage[export]{adjustbox}

\usepackage{hyperref}
\hypersetup{ colorlinks=true,
             citecolor=cyan,
             linkcolor=darkgray,
             urlcolor=teal,
             breaklinks=true,
             bookmarksnumbered=true,
             bookmarksopen=true,
}
\usepackage{url} 
\usepackage{cleveref}
\usepackage{tasks}
\usepackage{wrapfig}
\usepackage{paralist}
\usepackage{dsfont}

\DeclareMathOperator*{\argmax}{argmax}

\usepackage[numbers]{natbib}

\begin{document}

%\begin{verbatim}
%\end{verbatim}

\title{Efficient entanglement purification based on noise guessing decoding}

\author{André Roque}
\affiliation{Instituto Superior T\'{e}cnico, Universidade de Lisboa, Portugal}

\author{Diogo Cruz}
\affiliation{Instituto Superior T\'{e}cnico, Universidade de Lisboa, Portugal}
\affiliation{Instituto de Telecomunica\c{c}\~{o}es, Portugal}

\author{Francisco A. Monteiro}
\affiliation{Instituto de Telecomunica\c{c}\~{o}es, Portugal}
\affiliation{ISCTE - Instituto Universitário de Lisboa, Portugal}

\author{Bruno~C. Coutinho}
\affiliation{Instituto de Telecomunica\c{c}\~{o}es, Portugal}

%\date{\today}

\begin{abstract} 
In this paper, we propose a novel bipartite entanglement purification protocol built upon hashing and upon the guessing random additive noise decoding (GRAND) approach recently devised for classical error correction codes. Our protocol offers substantial advantages over existing hashing protocols, requiring fewer qubits for purification, achieving higher fidelities, and delivering better yields with reduced computational costs. We provide numerical and semi-analytical results to corroborate our findings and provide a detailed comparison with the hashing protocol of Bennet et al. Although that pioneering work devised performance bounds, it did not offer an explicit construction for implementation. The present work fills that gap, offering both an explicit and more efficient purification method. We demonstrate that our protocol is capable of purifying states with noise on the order of 10\% per Bell pair even with a small ensemble of 16 pairs. 
The work explores a measurement-based implementation of the protocol to address practical setups with noise. This work opens the path to practical and efficient entanglement purification using hashing-based methods with feasible computational costs.
Compared to the original hashing protocol, the proposed method can achieve the same fidelities with a number of initial resources up to one hundred times smaller. The proposed method seems well-fit for future quantum networks with a limited number of resources and entails a relatively low computational overhead.
\end{abstract}
\maketitle
\section{Introduction}

%\textcolor{blue}{Indeed, the noise-guessing decoding principle can be adapted on-the-fly to the time-variant statistics of the errors. This is common when using GRAND variants in classical communications, where, typically, the decoder knows the instantaneous signal-to-noise ratio (SNR) and that information can be used to modify the order of the error patterns to be tested. Knowing the structure of the used digital modulation, one can precompute the probability of certain error types for each SNR and sort them in decreasing probability order. Such tables can be stored for a number of fixed SNRs and selected each time the channel conditions change, as proposed in \cite{Chatzi_Mont_CommLett_2023}}.

%\textcolor{blue}{It is known that classical RLCs can reach Shannon's channel capacity and are easy do generate. Low-density parity-check (LDCP) codes are also capacity-achieving when decoded with the belief-propagation algorithm, but their construction imposes some constraints and not all $(n,k)$ pairs may be feasible. A fundamental characteristic of LDCP codes is the sparsity of the parity-check matrix. In a quantum setup using noise-guessing framework using a syndrome-based membership test, this sparsity induces a syndrome checking procedure that involves many fewer gates than the one using QRLCs. For this reason, for fault-tolerant syndrome computation, the use of LDPC codes may be preferable when using the proposed PGRAND, given their more limited exposure to gate errors.}

Entanglement is considered a valuable resource for numerous applications, spanning from secure communication and scalable quantum networks \cite{kimble2008quantum, caleffi2018quantum,PhysRevA.101.052315}, to precision measurement and timing, entanglement is a versatile resource for various quantum technologies \cite{horodecki2009quantum}, including quantum key distribution \cite{bennett2014quantum}, state teleportation \cite{bennett1993teleporting}, distributed quantum computation \cite{cirac1999distributed,dur2003entanglement} and distributed sensing  and metrology \cite{sekatski2017improved}. In experimental settings, notable progress has been made in establishing entanglement between different components, as evidenced by successful demonstrations in various systems, including trapped ions \cite{krutyanskiy2023entanglement, figgatt2019parallel,bruzewicz2019trappedion}, NV centers \cite{riedel2017deterministic,ruf2019optically}, neutral atoms \cite{kaufman2015entangling, madjarov2020highfidelity}, and superconducting circuits \cite{leung2019deterministic, flurin2015superconducting}. 

However, entanglement is susceptible to several detrimental factors. Foremost among these is decoherence, resulting from interactions with the environment or thermal fluctuations that cause a quantum system to lose its coherent superposition, ultimately leading to the breakdown of entanglement. Quantum errors, resulting from imperfect gates or measurements, can also disrupt the intricate entangled states \cite{horodecki2009quantum,roffe2019quantum,devitt2013quantum}. It is worth noting that the fidelity of the generated Bell pairs currently stands at approximately $10\%$, whereas the noise stemming from local gates and measurements is considerably lower, often falling below $1\%$.

Unless actively corrected, the accumulation of these error processes will inevitably destroy any initial entanglement. Purification protocols (PP) are a possible solution for quantum networks that has been extensively studied in the context of quantum repeaters \cite{Purification2023, munro2015quantum,duan2001longdistance} and is a crucial component of entanglement routing protocols \cite{santos2023shortest,bugalho2023distributing,coutinho2023entanglement}.

Bipartite entanglement purification encompasses a broad category of techniques aimed at acquiring high-fidelity copies of Bell states that are jointly held by two parties \cite{bennett1996mixedstate,bennett1996purification,deutsch1996quantum,chau2011practical, dur2003entanglement, krastanov2019optimized,miguel-ramiro2018efficient, dur2007entanglement, yan2023advances}. This is achieved by using local quantum operations and classical communication (LOCC) on an initially shared ensemble of entangled particles. Purification protocols are designed to operate on two or more ensembles of noisy entangled quantum states. The primary goal is to generate a reduced number of states with higher fidelity. Purification protocols can be repeatedly applied to the refined states in order to achieve states with even higher levels of purification.
Protocols such as the  \textit{IBM} protocol \cite{bennett1996purification,bennett1996mixedstate} or the \textit{Oxford} protocol\cite{deutsch1996quantum}, which fall into the category of recurrence protocols \cite{bennett1996purification, dur2007entanglement}, achieve this only probabilistically. Although the recursive application of these protocols leads to convergence towards maximally entangled states, their probabilistic nature requires multiple executions to successfully obtain a purified state. Since a single successful application only results in marginal improvements to the state's fidelity, the need to repeat these protocols several times to achieve sufficiently purified states leads to an exponential increase in the resources required. Another approach to purification is provided by hashing protocols \cite{bennett1996mixedstate, krastanov2019optimized, miguel-ramiro2018efficient, aschauer2005multiparticle}. Hashing protocols operate on a large ensemble of Bell pairs to deterministically produce a smaller number of purified pairs, whose fidelity can be arbitrarily high. However, traditional hashing protocols can be highly computationally demanding. Moreover, they often require hundreds of pairs to achieve purification and thousands of pairs to become efficient.

 In this work, we introduce a novel one-way bipartite purification protocol, inspired by the recently developed decoding technique for quantum error correction codes (QECCs) known as quantum random additive noise decoding (QGRAND)~\cite{cruz2023quantum,Chandra2023}. QGRAND was initially proposed by \cite{cruz2023quantum} for quantum random linear codes and latter has been used to decode other QECCs in \cite{Chandra2023}. QGRAND takes advantage of the noise statistics and provides an efficient decoding process. Employing a hashing-based approach, our protocol represents a substantial advancement over the hashing protocol \cite{bennett1996mixedstate}, with the distinguished advantages:  \begin{inparaenum}[i)]
    \item decreased qubit demand for purification,
    \item achievement of heightened fidelities for equal initial ensembles (maintaining equal qubit count and noise conditions),
    \item augmented yields for smaller ensembles,
    \item attainment of computationally feasible costs.
\end{inparaenum}

Hashing protocols, while efficient, encounter a significant practical limitation: their viability diminishes in real-world scenarios with noisy local operations and measurements. Even some minimal noise disrupts these protocols due to global operation requirements and the cumulative entropy increase from noisy operations. Building upon the concepts introduced by \cite{zwerger2014robustness,zwerger2016measurementbased}, we also put forward a measurement-based \cite{raussendorf2009measurementbased, raussendorf2001oneway} version of our protocol and provide numerical results regarding the protocol's potential noise tolerance and its impact on the performance.

This paper is organized as follows. In \Cref{sec:Background} we introduce a classical error decoding procedure along with its quantum equivalent and explain how a purification protocol can be obtained from a quantum error correction code.  In \Cref{sec:PGRAND} our purification protocol is presented and some results about its performance are provided. \Cref{sec:comparison} is devoted to establishing a comparison between our protocol and the hashing protocol. In \Cref{sec:measurement} we present a measurement-based version of our protocol before some conclusions being presented in \Cref{sec:conclusion}.

\section{Background}
\label{sec:Background}

\subsection{QGRAND}

While traditional decoding algorithms of classical error correction codes primarily concentrate on detecting error patterns through codeword analysis, a recent innovative proposal has surfaced, that focuses on ``decoding'' the noise; the algorithm was named GRAND (guessing random additive noise decoding) \cite{duffy2019capacityachieving}. This algorithm redirects the decoding process toward discovering the error pattern that affected a transmitted codeword rather than focusing on the codebook.

Consider a finite alphabet of symbols $\mathcal{A}$. Let $\mathcal{C}$ be a discrete communication channel with a block of $n$ input symbols denoted by $X^{n}$, which are subjected to a noise block $N^{n}$, also composed of $n$ symbols. Denoting the action of the channel on the input block by $\oplus$, the output block of $n$ symbols is $Y^{n} = X^{n} \oplus N^{n}$. 
Assuming that the channel's action is invertible, let $\ominus$ denote the inverse function, so that the inputs can be written as $X^{n} = Y^{n} \ominus N^{n}$. Suppose that this channel is employed for communication between a sender and a receiver, both of whom share a codebook $\mathcal{B}^{(n,M_k)}$ containing $M_k$ codewords, each of which is made of $n$ symbols taken from  the alphabet $\mathcal{A}$. Each particular random set of $n$ symbols forming a block $N_{i}^{n}$ is said to be a \textit{noise pattern}, which occurs with some known probability $P(N_i^{n})$, according to the noise statistics.
GRAND operates by successively testing noise patterns (in decreasing order of their probability) until finding one that generates a valid codeword, that is: $X^{n} = Y^{n} \ominus N_{i}^{n}$, such that $X^{n} \in \mathcal{B}^{(n,M_k)}$. When this is true, then $X^{n} = Y^{n} \ominus N_{i}^{n}$ is accepted as the decoded data. Otherwise, the algorithm moves to the next most likely error pattern.
It was shown in \cite{duffy2019capacityachieving} that this decoding procedure returns the maximum likelihood solution, i.e.,
\begin{equation}
    b_i^{(n,M_k)} = \argmax_{b_i^{(n,M_k)} \in \mathcal{B}^{(n,M_k)}} \{ P(N^{n} = Y^{n} \ominus b_i^{(n,M_k)})\}.
 \end{equation}
As mentioned, the probabilities for each noise sequence can be estimated using the specific noise statistics of the channel.

Quantum GRAND (QGRAND) is a recently proposed decoder that adapted the classical GRAND to a QEC setup \cite{cruz2023quantum}. It allows to decode quantum random linear codes (QRLC) with the distinguished advantages of maximizing the use of information about the noise while providing a simple encoding/decoding process. 
QGRAND was defined for stabilizer codes and leverages noise statistics to establish a recovery procedure similar to the one used by GRAND. 
Consider an initial setup of some $k$-qubit state that one wants to protect against some source of noise. The encoding is generated by randomly choosing two-qubit Clifford gates randomly selected and applied to the $n$ qubit system ($k$ initial qubits plus $n-k$ additional qubits in the state $  | 0 \rangle $). The target pairs of such gates are also randomly chosen, assuming that all-to-all connectivity between the $n$ qubits is possible. This encoding has been shown to be robust to depolarizing noise provided that the number of gates used is large enough.

QGRAND can be done implemented in slightly different ways. In the following, we describe the approach taken in the present work. Let $\mathcal{N}$ be the noise statistics of the error source considered. Since the circuit (composed of both the encoder and the stabilizers) is composed of Clifford gates, it is possible to classically simulate it in a efficient manner. By doing so, we can precompute a set $\mathcal{N}_{P} \subseteq \mathcal{N}$ of error patterns. A set $\mathcal{N}_J \subseteq \mathcal{N} \setminus \mathcal{N}_P$ of error patterns whose syndrome needs to be computed on the fly can also be considered, at the cost of some extra processing time. For each syndrome, only the most likely error pattern is stored, allowing one to establish a one-to-one correspondence between each syndrome and the set of correctable error patterns.
Such correspondences between syndromes and error patterns can be adjusted to different error models. Similarly, this is done in classical communications using GRAND: the testing order of the error patterns can be computed in advance for a number of different channel conditions and stored in tables. Those tables can be selected on the fly, according to varying channel conditions \cite{Chatzi_Mont_CommLett_2023}.
Once the syndrome has been extracted, a correction attempt is made. The correction process should be reversible, as the encoding, syndrome measurement, and decoding procedures can fail. In that case, the applied correction is reversed and another correction procedure is tried. This process is repeated until an error pattern that leads to a valid codeword is found.

\subsection{Equivalence between Quantum Error Correction and One-Way Purification Protocols}
\label{sec:EQ_EPP}

Both QEC and purification protocols (PP) are strategies for managing the challenges of noisy quantum communication that are intricately interconnected. This relation was first observed in the early exploration of entanglement purification in \cite{bennett1996mixedstate, aschauer2005quantum}, where a general methodology has been introduced to construct QECCs from one-way entanglement purification protocols. In cases where a code employs $n$ physical qubits to encode $k$ logical qubits ($k < n$), it becomes feasible to devise a purification protocol that makes use of $n$ copies of two-qubit states and produces $k$ pairs with enhanced fidelity.  The remaining $n-k$ pairs undergo measurement for revealing information concerning the obtained pairs \cite{matsumoto2003conversion,aschauer2005quantum}.

To understand the main idea, let us consider a noisy channel $\mathcal{C} $, established between two parties, A (Alice) and B (Bob), to be used for distributing  $n$ perfect Bell pairs. We denote the four Bell basis states as $ \{ | \phi^{\pm} \rangle = \frac{1}{\sqrt{2}}(|00 \rangle \pm | 11 \rangle), | \psi^{\pm} \rangle = \frac{1}{\sqrt{2}}(|01 \rangle \pm | 10 \rangle)  \} $
 and consider a Bell pair to be the state $ | \phi^{+} \rangle$, although any other of the 4 states could be considered. Although the construction can be generalized for any QEC code, we shall consider a stabilizer code for simplicity. For stabilizer codes, one can consider that the encoding and decoding operations consist of unitary operations $U_{\rm enc} = U$ and $U_{\rm dec} = U_{\rm enc}^{-1} =U^{-1}$. In order to obtain a purification protocol,  we start by noticing the fact that performing any unitary transformation (and this holds for any linear transformation) at B is equivalent to performing the transpose of that same transformation at A \cite{bennett1996mixedstate}, that is,

\begin{equation}
    U_{A}^{T} \otimes I_{B} \hspace{1mm} |\phi_{AB}^{+} \rangle^{\otimes n} = I_{A} \otimes U_{B} \hspace{1mm} |\phi^{+}_{AB} \rangle^{\otimes n}.
\label{eq:Bell_pair_equality}
\end{equation}

This allows us to perform encoding operations locally by A, and decoding to be performed locally by B. The channel affects the ensemble with some noise, which we shall assume to be a Pauli string $E$, and a mixed state
\begin{align}
    \rho &=\mathcal C(|\phi^{+}_{AB} \rangle^{\otimes n} )\\ 
    &=(1-p)(|\phi^{+}_{AB} \rangle \langle \phi^{+}_{AB} |)^{\otimes n} + pE(|\phi^{+}_{AB} \rangle \langle \phi^{+}_{AB} |)^{\otimes n}E^{\dagger}, 
\end{align}
is obtained. In order to purify $\rho$, Alice starts by applying the encoding on her qubits, projecting them into a $+1$ eigenspace defined by the code. Now Alice can perform measurements on $n-k$ of her qubits to obtain information about the ensemble $\rho$. Let $\{P_{i}\}$ be  the set of $2^{n-k}$ projectors defined by the code. Since each projector $P_{i}$ either commutes or anticommutes with any Pauli string, we have that 
\begin{equation}
    (P_{i} \otimes I)(I \otimes E) |\phi^{+} \rangle^{\otimes n} = \pm (I \otimes E)(P_{i} \otimes I) | \phi^{+} \rangle^{\otimes n}.
\end{equation}

Using \Cref{eq:Bell_pair_equality}, we obtain that
\begin{align} 
   (P_{i}^{2} \otimes I)(I \otimes E) |\phi^{+} \rangle^{\otimes n} = \pm (I \otimes E)(P_{i} \otimes P_{i}^{T}) | \phi^{+} \rangle^{\otimes n}.
\end{align}

Hence, every time Alice applies a projector $P_{i}$ on her qubits, Bob's qubits are projected through $P_{i}^{T}$.  After obtaining information about the measurements, Alice might restore her qubits to the $+1$ eigenspace defined by the encoding, although this is not strictly necessary if one ``adapts'' the measurement information to the projection that was performed. The measurement information is transmitted to Bob through a classical channel, allowing him to adjust the correction operation required to return his $n - k$ qubits to the code space. Finally, decoding operations are applied so that a state with higher fidelity is obtained. Thus, a purification protocol was obtained using only LOCC.

This construction can be summarized as follows:
\begin{enumerate}
\itemsep0em
    \item Generate $n$ copies of a maximally entangled bipartite state and distribute them between A and B.
    \item Apply locally on A the coding operation $U^{T}$ and Pauli measurements on $n - k$ of the $n$ qubits. The information about the measurements is sent classically to B.
    \item Extract the syndrome on B. Apply on B the correction operation determined from that measured syndrome.
    \item Apply the decoding operation $U^{-1}$ on B.    
\end{enumerate}

In the end, one obtains $k$ Bell pairs shared between Alice and Bob, while the remaining $n-k$ pairs are discarded. The $k$ purified pairs have higher fidelity than the initial $n$ shared pairs.

\subsection{Noise modeling}

In order to evaluate the performance of the proposed protocol, we model both coherent and decoherence errors in bipartite states through a depolarizing channel. A \textbf{depolarizing channel} $\mathcal{D}$ is a quantum channel that maps a quantum state onto a mixture of the original state and a maximally mixed state. If $\rho$ represents the density matrix of some state, the action of $\mathcal{D}$ on qubit $i$ of the state $\rho$ is naturally described by

\begin{equation}
    \mathcal{D}^{i}_{p}(\rho) = (1-p) \rho + \frac{p}{3}(X_{i}\rho X_{i} + Y_{i}\rho Y_{i} +Z_{i}\rho Z_{i}),
\end{equation}
where $p$ is the depolarization probability, that can be interpreted as an error probability, and $X,Y$ and $Z$ are the Pauli matrices. 

Using this channel to model noise requires the following assumptions: \begin{inparaenum}[i)]
    \item the qubit errors that affect one qubit are independent from the errors that affect the remaining system,
    \item single-qubit errors ($X$, $Y$ and $Z$) are equally likely and
    \item all qubits belonging to the same system have the same error probability $p$.
\end{inparaenum} Throughout this work, we will refer to noise of this form as local depolarizing noise (LDN). Applying a depolarizing channel to either the first or the second qubit of a Bell pair results in a state known as a Werner state, $W_{F}$, \cite{werner1989quantum} with fidelity  $F = \mathcal{F}(W_{F} , | \phi^{+} \rangle ) = \langle \phi^{+} | W_{F} | \phi^{+} \rangle = 1-p$. This enables us to express this state in terms of its fidelity:
\begin{align}
     \label{eq:Werner_Bell}
     \nonumber
W_{F} &= \mathcal{D}^{2}_{p}( | \phi^{+} \rangle \langle \phi^{+} |) = \mathcal{D}^{1}_{p}( | \phi^{+} \rangle \langle \phi^{+} |)\\
&= \frac{1-F}{3}I + \frac{4F-1}{3} | \phi^{+} \rangle \langle \phi^{+} |,
\end{align}
meaning that the obtained state can be interpreted as a classical mixture of the four Bell states. Notice that the qubit to which the channel is applied is irrelevant.

Werner states hold significance in the examination of random two-qubit states, as the conversion of any bipartite mixed state to a Werner state is attainable through an irreversible process termed ``twirl''. This procedure involves the application of independent and random rotations from the $\textit{SU}(2)$ group to each particle within the pair \cite{bennett1996mixedstate,dur2005standard}.

\subsection{Measurement Based Quantum Computation}

The measurement-based quantum computation (MBQC) \cite{raussendorf2009measurementbased, raussendorf2001oneway,raussendorf2002computational} is a model of quantum computation where computations are performed through measurements on an initial resource state, rather than using gates. It provides a universal framework for performing quantum computations utilizing only single-qubit measurements.

One of the most extensively studied and promising approaches in MBQC involves using as resource state a highly entangled state known as \textit{the cluster state}. These states can be generated in lattices of qubits through Ising type interactions \cite{raussendorf2001oneway,raussendorf2009measurementbased}.
This process generates a highly entangled state with the topology of a 2D square lattice, commonly referred to as a cluster state, as mentioned earlier. These cluster states are a specific type of state from a class of states that can be represented using graphs, known as graph states. To define a graph state, consider a graph $G = (V,E)$ with a finite set of vertices $V$ and edges $E \in V \times V$. The graph state associated with $G$ is defined as
\begin{equation}
    | G \rangle = \prod_{(a,b)\in E} CZ_{a,b} \left( \bigotimes_{a \in V} | + \rangle^{a} \right),
\end{equation}
with $CZ_{a,b}$ representing the controlled-$Z$ gate on qubits $a$ and $b$ and $| + \rangle = \frac{| 0 \rangle + | 1 \rangle }{\sqrt{2}}$. Some graph operations such as vertex deletion and edge addition or deletion have a physical correspondence with operations on the graph state. Keeping this in mind, two-dimensional cluster states can be defined as graph states obtained from a graph with the topology of a square lattice. By systematically choosing the measurement sequence and bases, any arbitrary quantum circuit can, in principle, be simulated utilizing only single-qubit measurements on the cluster state \cite{raussendorf2009measurementbased}.

This paradigm of quantum computation is especially relevant for the development of fault-tolerant quantum computation \cite{nielsen2005faulttolerant,dawson2006noise,brown2020universal,gimeno-segovia2019deterministic}. Moreover, experimental realizations of multiparty purification protocols in optical lattices, aimed at improving the fidelity of cluster states, have already been presented \cite{dur2003multiparticle, aschauer2005multiparticle}.

\section{Proposed Protocol}
\label{sec:PGRAND}
\subsection{Protocol Overview}

\begin{figure*}[!ht]
\centering
\includegraphics[width=\textwidth]{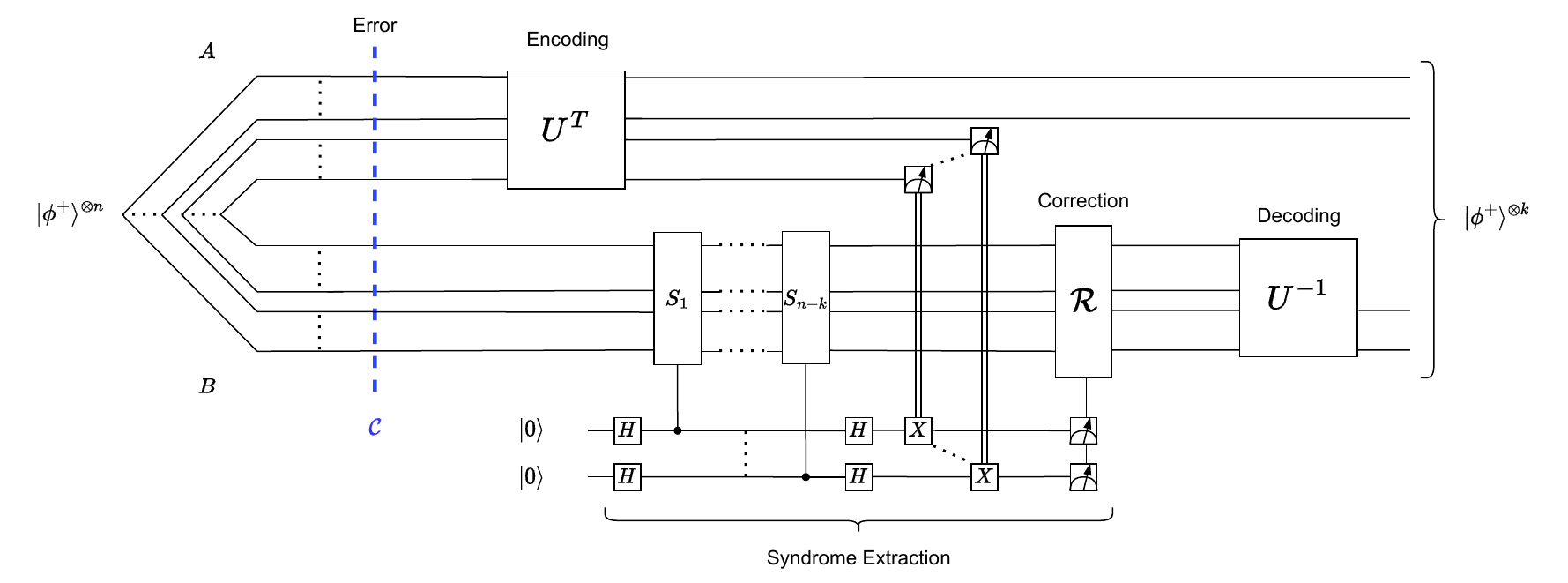}
\caption[Quantum circuit for PGRAND applied to $n$ Bell pairs distributed by parties A and B in order to obtain $k$ purified pairs.]{Quantum circuit for the PGRAND applied to $n$ Bell pairs distributed between Alice (A) and Bob (B) in order to obtain $ k$ purified pairs. The error is described by a quantum channel $\mathcal{C}$. In the circuit, the measurements performed by Alice are depicted changing the ancilla qubits, but in reality only a virtual syndrome update needs do be done. The same goes to the recovery procedure: if it consists only of Pauli strings, then $\mathcal{R}$ can be performed virtually (in software), without the need to apply extra gates.}
\label{fig:PGRAND_scheme}
\end{figure*}

Following the construction proposed in \Cref{sec:EQ_EPP}, we now present a purification protocol developed through an adaptation of QGRAND, which we will refer to as Purification GRAND (PGRAND). Let us assume an initial setup of $n$ Bell pairs shared between Alice and Bob through a noisy quantum channel $\mathcal{C}$. At the end of the protocol procedures, both parties will share $k$ Bell pairs with higher fidelity. The protocol is depicted in \Cref{fig:PGRAND_scheme} and can be summarized by the following steps:

\begin{enumerate}
\itemsep0em
    \item Apply a random encoding on Alice's $n$ qubits. Share the encoding information with Bob.
    \item Perform single-qubit measurements in the computational basis on $n-k$ of Alice's $n$ qubits, followed by the classical transmission of the measurement outcomes to Bob.
    \item With the information about the encoding, simulate the stabilizer circuit and determine the stabilizers for the qubits which were initially entangled with the measured ones. 
    \item Using the stabilizer information, compute the syndromes associated with the most likely errors. Construct a lookup table (LUT) containing the most likely errors associated to each syndrome.
    \item Measure the syndrome using all the determined $n-k$ stabilizers.
    \item Update the syndrome obtained with the information sent by Alice.
    \item Use a noise guessing approach to identify the error pattern and apply the recovery procedure on the corrupted state.
    \item Apply the decoding at B.  Discard the measured pairs and keep the $k$ remaining ones.
  
\end{enumerate}

\subsubsection{Encoding}

The protocol starts by applying an encoding on Alice's qubits, composed of randomly selected 2-qubit Clifford gates. To do so, one can generate a random tensor product of Clifford gates, transpose it, and apply it to the initial state. The information about the original tensor product has to be sent classically to Bob. Alternatively, both parties could previously agree on the random encoding to be used. To obtain a highly performant code, $0.14n\log^{2}(n)$ Clifford gates are required on average \cite{cruz2023quantum}. As most of the gates can be applied in parallel, the encoding has a circuit depth of $\order{\log^3 (n)}$. We remark that, while the encoding should be generated randomly, the same encoding can be applied for subsequent uses, provided that the total number of pairs remains the same. By reusing the code, we may skip steps 3 and 4, which are the most computationally heavy of the entire protocol. This reuse significantly reduces the computational overhead associated with generating new LUTs for each round. Additionally, even when there are variations in the noise parameters, the precomputed LUTs can remain effective if these variations are minimal. For more complex changes in error rates across different qubits, the existing LUTs may still be employed, albeit with potentially suboptimal decoding performance. Precomputing LUTs for different anticipated noise channels or employing robust codes that are adaptable to on-the-fly modifications can further mitigate the computational demands and ensure efficient error correction. This flexibility and adaptability in the LUT management underscore the feasibility of our approach in practical quantum error correction scenarios.

\subsubsection{Determining the stabilizers}

After the qubits are generated and subjected to noise, the process of gathering information about the state is limited to local operations. To overcome that issue, we can still make use of local stabilizers on Bob's qubits, but it is required that the respective entangled qubits on Alice's side be measured. After some initial state $| \psi \rangle_{I}$ has been encoded with an unitary transformation $U$, the stabilizers of the encoded state $| \psi \rangle_{F} = U | \psi \rangle_{I}$ can be obtained by evolving the initial stabilizers through the encoding, that is, if $S_{I}$ is a stabilizer of the initial state, then $S_{F} = U S_{I} U^{-1}$ is a stabilizer of the final encoded state since

\begin{align}
    &S_{F} | \psi \rangle_{F} =  (U S_{I} U^{-1}) U | \psi \rangle_{I} =  U S_{I} | \psi \rangle_{I} = \\ \nonumber &U | \psi \rangle_{I} = | \psi \rangle_{F}.
\end{align}

Although the proposed protocol performs the measurements after the encoding operation, given \Cref{eq:Bell_pair_equality} it is possible to assume that, for those qubits that are being measured, the measurements on Alice's side are performed before the encoding operation on Bob's side. Hence, we can focus our attention in evolving the stabilizers of the measured states. If the initial state is a Bell pair, the operators $S_{M = 0} = I_{A} \otimes Z_{B}$ and $S_{M = 1} =I_{A} \otimes (-Z)_{B}$ are able to stabilize locally on Bob's side the states obtained after a $Z$-measurement of a Bell pair, $|00 \rangle_{AB}$ and $|11 \rangle_{AB}$, respectively. Hence, each stabilizer can be evolved through $U^{T}$, that is, we obtain the stabilizers $S'_{M = 0} = I_{A} \otimes ((U^{T})^{-1}ZU^{T})_{B}$ and  $S'_{M = 1} = I_{A} \otimes (-(U^{T})^{-1}ZU^{T})_{B}$. Since $S'_{M = 1} = - S'_{M = 0}$, changing the stabilizers according to the measurement is equivalent to performing a classical bit-XOR operation of the syndrome with the measurement result.

According to the Gottesman-Knill theorem \cite{gottesman1998heisenberg,aaronson2004improved}, for a quantum circuit of $n$ qubits, updating the stabilizer specification requires $\mathcal{O}(n)$ time for each Clifford gate in the circuit. The encoding operation made by $U$ is a stabilizer circuit, and thus it has depth $\order{\log^3(n)}$, determining the stabilizers requires $\mathcal{O}(n\log^{3}(n))$ time.

\subsubsection{Noise guessing and correction}\label{sec:noise_correction}

Bob can efficiently simulate the encoding circuit and build a parity check matrix to map each error pattern to the respective syndrome. With the noise statistics, Bob can determine the most likely error associated with a syndrome, so that whenever that syndrome is measured, the recovery procedure is done such that the most likely error is selected and corrected.
 
Since the possible number of error patterns increases exponentially on the number of qubits for most of the error models, Bob might only be able to precompute the syndromes associated with a fraction of the total possible errors. The number of syndromes that are computed should be chosen according to the initial fidelity of the pairs, the number $n$ of qubits and the required final fidelity. This approach will save resources and make the computation possible at the cost of a decrease in performance. 
The computation of the syndrome of each error pattern involves multiplying a $2n$-bit array representing the error with a $(2n)\times(n-k)$ matrix representing the stabilizers, requiring  $\mathcal{O}(n^{\omega})$ time. In practice, this procedure can be efficiently parallelized. When aiming to correct the $N$ error patterns with a weight up to $t$, we may take advantage of previously computed syndromes for lower weight errors to achieve a $\mathcal{O}(Nn)$ computation cost in obtaining the syndrome table.  

It is important to note that this step is the primary limiting factor in what regards the classical computational cost, just like in the case of the hashing protocol \cite{bennett1996mixedstate,chau2011practical}, with the circuit simulation playing a negligible role in the overall cost.

After mapping each error pattern with a syndrome, only the most likely error pattern associated to each syndrome needs to be stored, and the remaining can be discarded.
The number of possible error patterns is $N$ and $S = 2^{(n-k)}$ is the number of available syndromes. Whenever a syndrome is extracted, all that is needed to identify the recovery procedure is a linear search on a table of at most size $T=\min\qty{N, S}$, which requires $\mathcal{O}(Tn)$ space and time to store the syndromes and perform a search.

\subsubsection{Decoding}

As outlined in the protocol's presentation, the decoding entails reversing the encoding operation and discarding the measured qubits. Nevertheless, since only Clifford gates are used in the decoding, a virtual execution prior to the recovery operation is also feasible. In that case, all that needs to be done is to make the appropriate adjustments to the recovery operation. Both the decoding (except for the discarding of pairs) and the correction involve Clifford gates, so all of this can be performed classically through software \cite{paler2014softwarebased}.

Overall, for a single PGRAND application that uses $n$ Bell pairs to obtain $k$, the overhead per pair, that is, the number of qubits that are spent in the use of the protocol, is given by $3(n/k-1)$ (due to the $2(n-k)$ qubits that are required to be measured and the $n-k$ qubits that are used as ancillas). For the syndrome extraction, each of the $n-k$ stabilizer applications is composed by, on average, $3n/4$ CNOT gates and $7n/2$ single-qubit gates, requiring a total of $\order{n^2}$ gates.

\subsection{Results}

Consider a setup of Werner states with initial fidelity $F_{i} = 1-p$, with $p$ being the parameter of a depolarizing channel. Define $\mathcal{W}: \mathcal{E} \rightarrow \mathbb{N}$  to be the map between each error pattern $E_{i} \in \mathcal{E}$ and its weight. 
Let $N_{w}$ be the number of possible error patterns with weight $w$ and $N_{\leq w}$ the number of possible error patterns with equal or less weight than $w$. For depolarizing noise these quantities are given by
\begin{equation}
    N_{w} = \binom{n}{w} 3^{w} , \quad N_{\leq w} = \sum^{w}_{i=0} N_i.
\end{equation}

Let $f_{\textrm{Bin}(n,p)}(w)$ be the binomial probability mass function. For each error pattern $E_{i}$, we have that
\begin{align}
    &P(\{E_{i}\} | \mathcal{W}(E_{i})=w) = \frac{f_{\textrm{Bin}(n,p)}(w)}{N_{w}} \\ \nonumber &= \qty(\frac{p}{3})^{w}(1-p)^{n-w},
\end{align}
from which we immediately conclude that $P(\{E_{i}\})\geq P(\{E_{j}\}) \Rightarrow \mathcal{W}(E_{i})\leq \mathcal{W}(E_{j})$ for any $E_{i},E_{j} \in \mathcal{P}^{n}$ (as long as $p<3/4$).  

The approach of correcting the most likely errors in this noise scenario prioritizes the correction of lower-weight errors. That is, for a set of error patterns that share the same syndrome, the protocol considers the error that has occurred to be the error pattern with the lowest weight. 

Under the assumption that any correctly identified error can be completely corrected, it is useful for the analysis of the protocol to define the correctable fraction of weight $w$ as
\begin{equation}
    f_{w} = \frac{|\{E_{i}: \mathcal{W}(E_{i})=w \; , \; E_{i} \; \textrm{is correctable} \}|}{|\{E_{i}: \mathcal{W}(E_{i})=w \}|}.
\end{equation}

Then, if degenerate scenarios are disregarded, the average correctable fraction is approximately \cite{cruz2023quantum}
\begin{equation}
     \langle f_{w} \rangle  \simeq \frac{S}{N_{w}}e^{- \frac{N_{\leq(w-1)}}{S}}(1-e^{-\frac{N_{w}}{S}}) \quad \textrm{for $S \gg 1$},
     \label{eq:theoretical_fraction}
\end{equation}
Note that computing the syndromes associated with every possible error pattern has a high computational cost and it might be impossible even for relatively small values of $n$. Thus we introduce a threshold  parameter $t\leq n$ that defines the maximum weight of the error patterns for which the syndromes are computed, implying that $f_{w} = 0, \quad \forall_{ w>t}$. Now we can compute probability of error $p_{e}$ of the protocol as follows

\begin{equation}
    \label{eq:theoretical_bler}
     p_{e} = 1 - \sum_{i=0}^{n} \langle f_{i} \rangle f_{\textrm{Bin}(n,p)}(i) = 1 - \sum_{i=0}^{t} \langle f_{i} \rangle f_{\textrm{Bin}(n,p)}(i).
\end{equation}
This provides a useful lower bound for the achieved average fidelity, $\langle F_{a} \rangle$, of the output pairs, as explained in \Cref{app:bler}. Therefore, we say that the protocol achieves purification if one obtains that $ \langle F_{a} \rangle \geq 1-p_{e}>F_{i}$. Note that, as it is defined, the probability of error corresponds to the probability of not correctly identifying an error pattern. This means that, in the limit situation, where no syndrome is extracted (and consequently no correction is applied to the ensemble), one obtains  $p_{e} = 1 - f_{\textrm{Bin}(n,p)}(0)$, while in practice the fidelity achieved is equal to the initial fidelity.

The efficiency of a protocol is determined by its yield, which is normally defined as the ratio between the number of purified pairs with fidelity arbitrarily close to unity and the initial number of pairs required by the protocol , that is, for a protocol that uses $n$ Bell pairs to produce $k$ Bell pairs with fidelity $F\geq 1 - \epsilon \quad \forall_{\epsilon > 0}$, the yield is defined as $\lim_{n \rightarrow \infty} \frac{k}{n}$\cite{bennett1996purification, dur2007entanglement}.  Such a definition would imply that many protocols would have a zero yield, including the hashing protocol when employed for finite-sized ensembles. Therefore, we shall consider a relaxed version of this definition, by defining the yield $Y$ simply as the ratio between the number of output and the number of input Bell pairs in analogy to the rate in the context of QECCs, as it is a common pratice in the literature \cite{krastanov2019optimized,miguel-ramiro2018efficient}.

\begin{figure}[t]
    \centering
    \includegraphics[width=0.9\columnwidth]{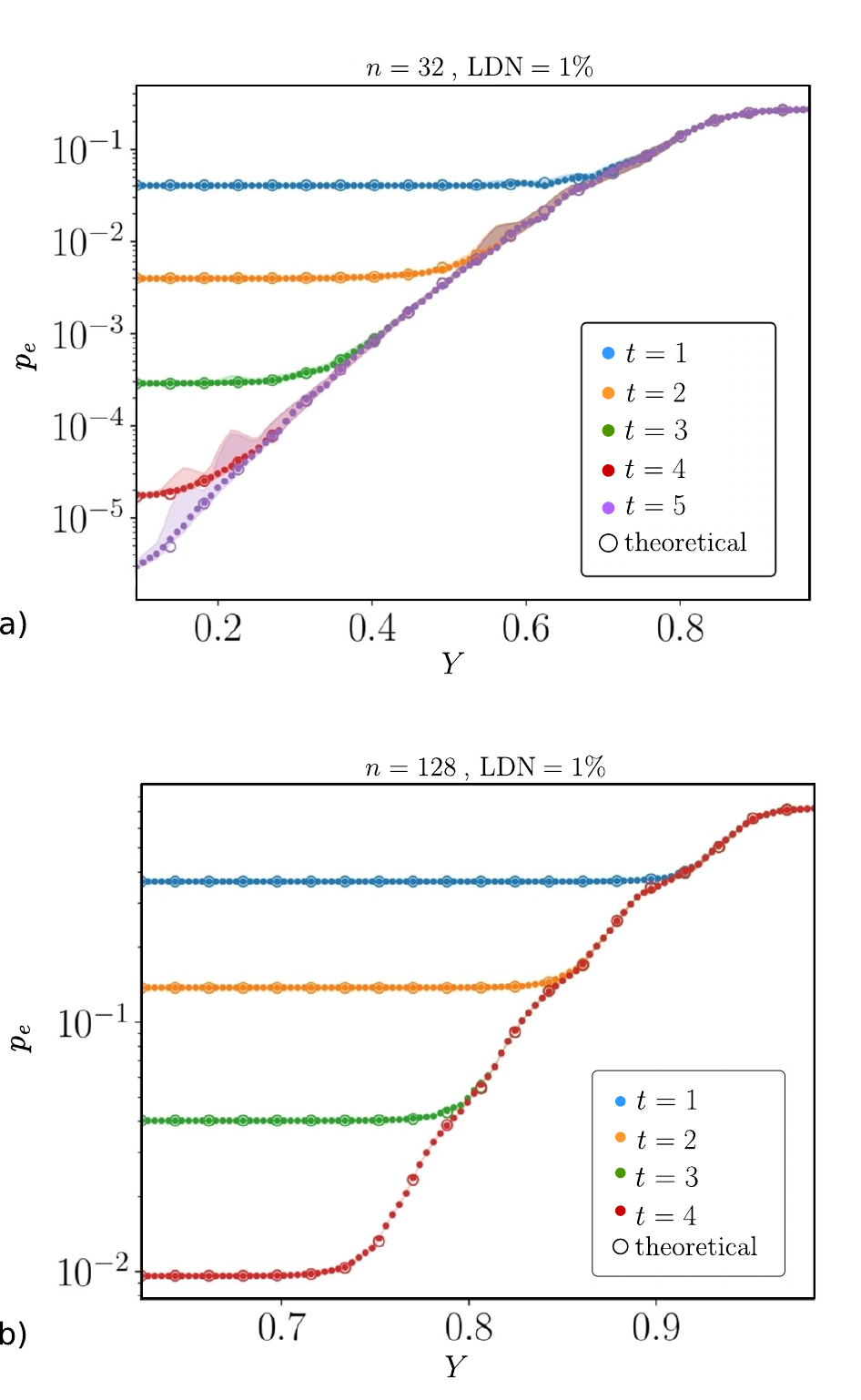}
    \caption{Simulation results of the protocol probability of error $p_{e}$ as a function of the yield, for a) 32 and b) 128 Bell pairs, assuming $1\%$ of LDN. While on plot a) we simulated the procedure for errors with a weight up to $t = 5$, on b) we restricted to $t = 4$. The larger white dots represent the values obtained by combining expressions \eqref{eq:theoretical_fraction}, and \eqref{eq:theoretical_bler}.
    The simulation results show that the probability of error closely agrees with the values obtained via the theoretical expression when using a relatively large number of gates for the encoding (120 gates for plot a) and 1000 for plot b)).}
    \label{fig:bler}
\end{figure}

To assess the accuracy of the results obtained using \Cref{eq:theoretical_fraction}, the random encoding was simulated by generating a random parity check matrix, and the syndromes for each considered error pattern were computed using the techniques explained in Ref.~\cite{gottesman1998heisenberg}. This was performed considering 32 and 128 Bell pairs, affected by 1\% LDN. The results are shown in \Cref{fig:bler}. These results were obtained for each value by performing twenty Monte Carlo simulations for each yield value considered. In the figure, each data point represents the average result of these simulations. The plots in the figure illustrate how the computational cost, represented by the parameter $t$, constrains the purification capabilities of the protocol. The presence of a horizontal line on each curve signifies the existence of a saturation point. At this point, the majority of the analyzed error patterns are successfully mapped to a syndrome, and further sacrificing of pairs (i.e., utilizing lower yields) does not translate into any additional advantage, meaning that in scenarios characterized by high levels of noise and/or a large number of pairs, the computational cost emerges as a limiting factor in achieving purification.

Given the computational complexity and hardware constraints associated with conducting these extensive simulations, the subsequent findings in this section are derived from their analytical approximations. The numerical results obtained using these expressions for various values of $n$, $p$, and $t$ can be found in \Cref{app:additional}
 
The approach of correcting the most likely errors implies that, in order to achieve purification, at least all the error patterns with weight up to $np$ must be considered by the protocol.

The capacity (i.e., the highest rate or yield one can possibly obtain using a quantum error correction code and PP) of the depolarizing channel for non-degenerate stabilizer codes is upper bounded by
\begin{equation}
\label{eq:DC_capacity}
\frac{k}{n} < 1 - p\log_{2}(3) - H(p),
\end{equation}
where $H(p) = -p \log_{2}(p)-(1-p) \log_{2}(1-p) $ is the binary entropy function \cite{ekert1996quantum}. This bound is known as the quantum Hamming bound, and it  can be exceeded by
degenerate codes \cite{divincenzo1998quantumchannel}. Although the capacity of the depolarizing channel has already been determined \cite{holevo1998capacity,king2003capacity}, this bound provides a more manageable expression for evaluating the performance of a stabilizer code in a setup that involves a large number of channel uses.

\begin{figure}[t]
    \centering
    \includegraphics[width=\columnwidth]{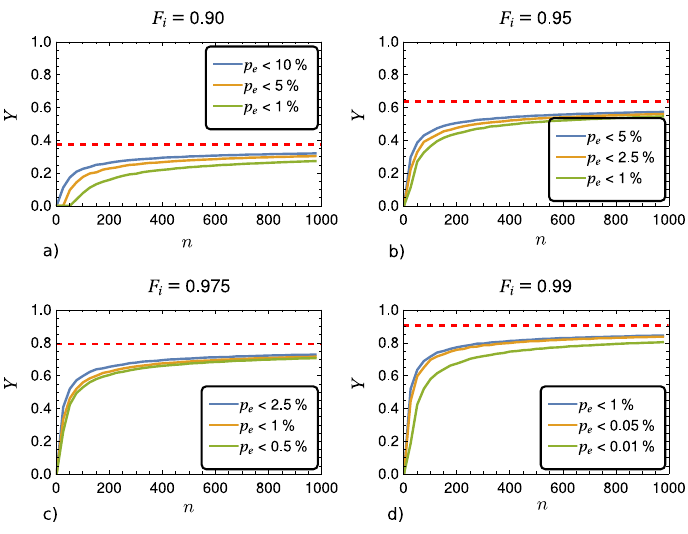}
    \caption[Maximum achievable yield to purify $n$ Bell pairs with an initial fidelity $F_{i}$]{Maximum achievable yield to purify $n$ Bell pairs with an initial fidelity  $F_{i}$ with a probability of error inferior to $p_{e}$. The initial fidelity is equal to a) $F_{i}=0.90$, b) $F_{i}=0.95$, c) $F_{i}=0.975$ and d) $F_{i}=0.99$. For each graph, the blue line represents the maximum yield at which purification is achievable. The dashed red line represents the bound given by \Cref{eq:DC_capacity}. Applying the protocol to ensembles with more than one hundred pairs allows for a substantial increase in the efficiency of the protocol.}
    \label{fig:maximum_yield}
\end{figure}

Using similar arguments as the ones in \cite{gottesman1997stabilizer}, one can prove that the yield of our protocol is upper bounded by this bound.
However, it is worth evaluating how close the yield requirements are to that bound. Assuming an initial fidelity $F_{i}$, \Cref{fig:maximum_yield} illustrates the maximum yield required to achieve an error probability below a specified threshold. These results are derived without imposing any constraints on the weight of the errors considered. Thus, the number of available syndromes emerges as the sole limiting factor influencing the outcomes.

The most significant observation from \cref{fig:maximum_yield} is the initial growth in the required yield as the number of pairs increases up to a few hundred. Notably, by employing the protocol on ensembles with a few more qubits, much superior efficiencies can be obtained. However, using more than five hundred pairs leads to only marginal changes in the yield required to obtain the same error probabilities, since in this regime the yield is already close to the capacity of the channel. This observation may have critical implications for the realistic implementation of the protocol, as the challenge of generating and manipulating larger qubit ensembles could pose a substantial obstacle. Another noteworthy aspect is the minimal difference in the required yields for achieving pairs with varying error probabilities. This is evident by the proximity between the curves in each plot, indicating that a much more purified state can be obtained by slightly sacrificing the protocol's yield.

\subsection{Computational limitations}

\begin{figure}[t]
    \centering
    \includegraphics[width=0.5\textwidth]{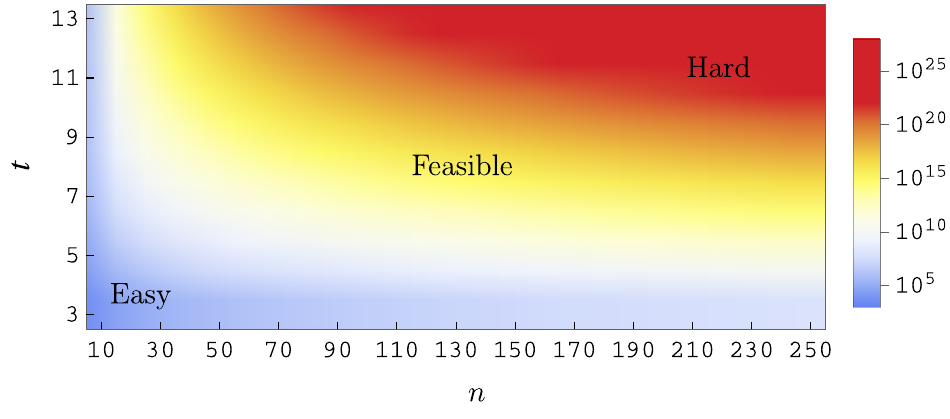}
    \caption{Contour plot of the number of possible error patterns with weight equal or inferior to $t$ on an ensemble of $n$ Bell pairs. Here it is possible to see that the number of error patterns increases exponentially with their weight, but sub-exponentially with the size of the ensemble. The yellowish zone marks the frontier between what is computationally feasible with the actual standards.}
    \label{fig:number_patterns}
\end{figure}

In some situations, it may prove infeasible to account for error patterns that are extremely unlikely, such as those with very high weight, when the noise model is a Pauli channel. \Cref{fig:number_patterns} showcases the regimes that are more realistic to consider. 

\begin{figure}[thp]
   \includegraphics[width= \columnwidth]{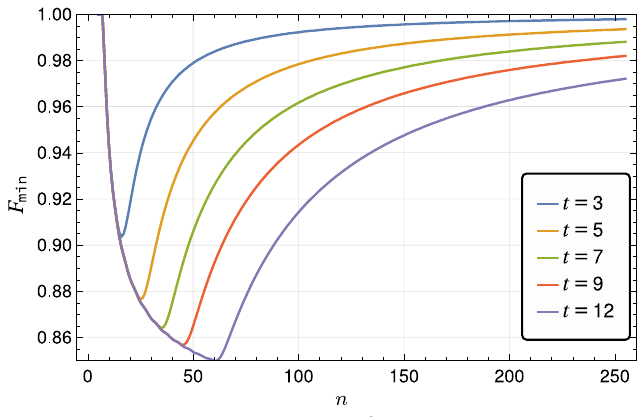}
   \caption{Minimum initial fidelity $F_{\textrm{min}}$ required to achieve purification as a function of the number of pairs in Werner form when attempting to correct errors with a weight up to $t \in \{3,5,7,9,12\}$. These results were obtained by considering that only one purified Bell pair is obtained $n$-to-one. If no constraints were imposed on $t$, the value of $F_{\text{min}}$ would strictly decrease with the number of pairs. However,  if one limits  the number of precomputed syndromes, a saturation point is reached.}    
    \label{fig:minimum_initial_fidelity}
\end{figure}

As we limit the correction process to only account for error patterns with weight at most $t$, PGRAND's performance may be degraded. In \Cref{fig:minimum_initial_fidelity} it is possible to see the minimum value of fidelity needed to achieve purification, $F_{\textrm{min}}$, according to the number of pairs and the constraint imposed on the value of $t$. If no constraints were imposed, this value is expected to solely decrease with the number of pairs, since in this scenario the only limiting factor is the number of available syndromes.

However, following \cref{sec:noise_correction}, if one limits the number of error patterns considered, a saturation point is reached. In this scenario, the number of possible syndromes surpasses the number of considered error patterns, and no advantage is obtained by having more syndromes. As the number of pairs increases, errors with greater weight become more likely, causing an increase in the error probability and the value of $F_{\textrm{min}}$.

%\resizebox{0.5\textwidth}{!}{
\begin{table*}[!htpb]
\centering
\begin{tabular}{|c|c|c|c|c|c|c|c|c|}
\hline
$t$       & 6      & 7      & 8      & 9      & 10     & 11     & 12     \\ \hline
$n$       & 30     & 35     & 40     & 45     & 50     & 56     & 61     \\ \hline
$F_{\textrm{min}}$ & 0.8695 & 0.8642 & 0.8601 & 0.8578 & 0.8542 & 0.8512 & 0.8499 \\ \hline
\end{tabular}
\caption{Minimum fidelity to achieve purification and respective number of pairs required as a function of the parameter $t$.  }
\label{tab:minimum_fidelities}
\end{table*}
%}
It is also possible to determine a bound for the value of $F_{\textrm{min}}$ regardless of the number of pairs, following a similar reasoning as the one used in Ref.~\cite{bennett1996mixedstate}. Indeed, given that the yield of the protocol is bounded by \Cref{eq:DC_capacity}, a positive yield can only be obtained as long as $p >0.1893$, that is, $F_{\textrm{min}} >0.8107$.

\section{Comparison of approaches}
\label{sec:comparison}

\subsection{The Hashing Protocol}
\label{sec:comparison_hashing}
A similar approach to purification is performed by the Hashing protocol. The Hashing protocol \cite{bennett1996mixedstate} is a one-way entanglement purification protocol that operates on a large ensemble of noisy $ n$ entangled pairs. Representing by $W$ the density matrix of the initial ensemble, the hashing protocol is capable of distilling a smaller number of pairs $k \approx n((1-S(W))$, where $S(W)$ is the Von Neumann entropy, with fidelity arbitrarily close to unity in the asymptotic regime where $n \rightarrow \infty$.

The protocol consists of $n-k$ rounds where, at each round one of the parties performs some unitary operations and one of the qubits is measured and sacrificed to reveal information about the system, with each measurement revealing one bit about the unmeasured pairs. The result of the measurement is sent classically to the other party, which by measuring the corresponding qubit obtains information about the parity of the ensemble.

For a small parameter $\delta >0$, the protocol attempts to correct  errors that belong
 to the following set
\begin{equation}
\label{eq:ts}
    T_{\delta} = \qty{ E_{i}: \qty|-\frac{1}{n} \log_{2}(P(\{E_{i}\}) - S(W) |< \delta  }.
\end{equation}

While the original intention of the hashing protocol was to operate with an asymptotically large number of Bell pairs, our interest is to assess its performance with a finite (and possibly small) ensemble.
For this reason, we consider the yield of the hashing protocol with $k$ rounds to be given by $k/n$.

Nonetheless, we present a comparison of performance between PGRAND and this protocol. The first challenge of establishing a fair comparison lies in the absence of a straightforward expression for the fidelity of the states produced by the hashing protocol. To enable the comparison, we rely on a bound derived in Ref.~\cite{zwerger2018longrange} concerning the average fidelity achieved by the hashing protocol when applied to Werner states with fidelity parameter $F_{i}$. The bounding expression is as follows:

\begin{align}
\label{eq:fidelity_bound_hashing}
\nonumber
    \langle F_{a} \rangle & \geq 1 - 2e^{\left\{ \frac{-n}{a(F_{i})} \left[(g(F_{i})+\delta)\ln{\left(1+\frac{\delta}{g(F_{i})}\right)-\delta}
 \right]  \right\}} \\ \nonumber
 &- 2^{-n[S(F_{i})+\delta]-(n-k)}, \\
 \nonumber \textrm{where} \\ \nonumber
&S(F) = \\ \nonumber &-F \log_{2}(F) - (1-F)\log_{2}\left(\frac{1-F}{3}\right), \\ \nonumber
&a(F) = \\ \nonumber &  \log_{2}\left(\frac{1-F}{3}\right)|+ S(F),\\ \nonumber
&g(F) =\\\nonumber &\frac{ F\log_{2}^2(F) + (1-F)\log_{2}^{2}\left(\frac{1-F}{3}\right)-S^{2}(F)}{a(F)}.\\
\end{align}

This bound was established by bounding the probability of failure of the hashing protocol 

% \footnote{This bound was obtained by bounding the probability of having an error pattern that falls outside of the considered set of errors through Bennet's inequality \cite{bennett1962probability} and the probability that two error patterns agree on the parity measurement. Despite being a bound, we can still use for comparison with our protocol, since the differences between the real values and this bound necessarily decrease exponentially in the number of used pairs, having a negligible difference for the type of results that are presented in this section.} using a similar reasoning to the one presented in \Cref{eq:bler}, in \cref{app:bler}. 

\begin{figure}[t]
    \centering
    \includegraphics[width=\columnwidth]{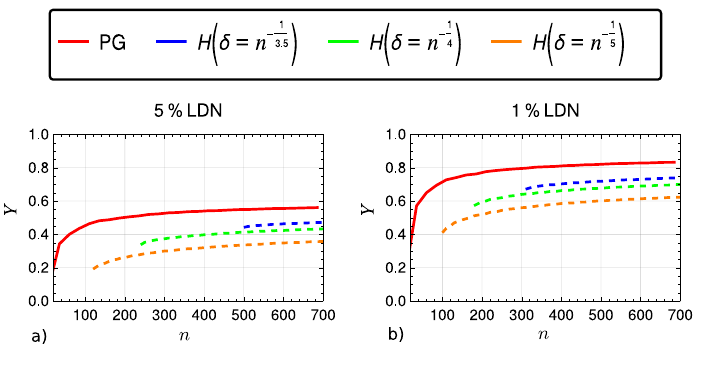}
    \caption[Minimum required fidelity $F_{i}$ for purification to be possible]{Average fidelity achieved $\langle F_{a} \rangle $ of the output pairs from the hashing protocol and PGRAND as a function of the yield, for 128 (a) and 256 (b) pairs. For $\delta = n^{-\frac{1}{3.5}}$ the hashing protocol is unable to achieve purification, meaning that considering lower values of $\delta$ is pointless. For the PGRAND curve, no constraints were imposed on the value of $t$. It is possible to observe the trade-off between the yield and output fidelity that comes from choosing different values for the parameter $\delta$ of the hashing protocol.  }    
    \label{fig:yield_fidelity_hashing}
\end{figure}

The parameter $\delta$ presents a challenge to our comparison, as its smaller values prompt the hashing protocol to attempt to correct a larger number of error patterns, as stipulated by \cref{eq:ts}. However, if the yield is not sufficiently small, an increase in $\delta$ leads to a reduction in the fidelity of the output pairs ( see \Cref{app:ts}). This trade-off is evident in \Cref{fig:yield_fidelity_hashing}, where we compare the fidelity achieved by PGRAND (without imposing a restriction on the number of corrected errors) to that achieved by the hashing protocol, for similar values of $\delta$ as the ones used in other references \cite{bennett1996mixedstate, zwerger2018longrange,miguel-ramiro2018efficient}.

Nevertheless, it is clear that PGRAND outperforms the hashing protocol, particularly when 128 pairs are used. For $\delta > n^{-\frac{1}{3.5}}$ the hashing protocol fails to achieve purification for the number of pairs considered.

\begin{figure}[tb]
    \includegraphics[width=1\columnwidth, left]{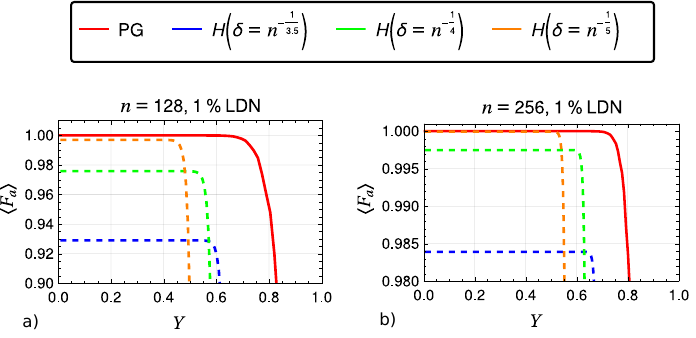}
    \caption[Yield required to achieve purification for the hashing protocol and PGRAND]{Yield required to achieve purification as a function of the number of pairs of the initial ensemble for the hashing protocol and PGRAND, for 1\% (a) and 5\% (b) of LDN. For the PGRAND curve, no constraints were imposed on the value of $t$. For the hashing protocol, increasing the number of pairs allows the choice of a smaller $\delta$. As the number of pairs increases, the differences between the efficiencies of the protocols become smaller.}
    \label{fig:yield_pairs_hashing}
\end{figure}

Next, we compare how the initial pair count in the ensemble changes the yield required for purification ($F_{a} \geq 1- p$). The results in \Cref{fig:yield_pairs_hashing} show us that the hashing protocol demands a larger ensemble of pairs. The increase of the number of pairs enables the use of a greater value for the parameter $\delta$, leading to higher yields. It becomes clear that as $n$ increases the hashing protocol performance approaches the performance of our protocol. In the considered noise scenarios, using higher values of $\delta$ would lead to similar yields for both protocols. 

To evaluate the minimum number of pairs required, denoted as $n_\textrm{min}$, for achieving purification, the protocols were examined under the condition where only one purified pair is generated. Specifically, this corresponds to the case where $k = 1$, enabling the establishment of a limiting scenario. For the PGRAND evaluation, no constraints were imposed on the value of $t$ (which corresponds to considering $t = n$). 

\begin{figure}[t]
\centering
   \includegraphics[width=\columnwidth]{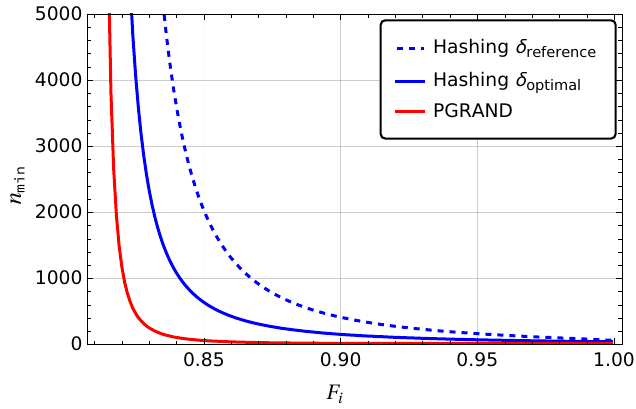}
   \caption{Number of pairs in Werner form with fidelity $F_{i}$ required to achieve purification. While in both protocols there is an exponential increase in the number of required pairs with the amount of noise, the requirements of our protocol are substantially lower than the ones of the hashing protocol.}
   \label{fig:number_pairs} 
\end{figure}

When setting $k = 1$ for the hashing protocol, it is common in the literature to use $\delta_{\textrm{reference}} = \frac{1}{2} \left(\frac{n-1}{n} - S(F)\right)$ \cite{zwerger2018longrange,miguel-ramiro2018efficient}. However, this choice of $\delta$ is suboptimal, as demonstrated in \Cref{fig:number_pairs}, where we compare the values of $n_\textrm{min}$ for our protocol and the hashing protocol. Therefore, we include a comparison with the results obtained by the hashing protocol when selecting an $\delta_{\textrm{optimal}}$ that maximizes the fidelity for each considered fidelity value $F_{i}$. The results at some notable points of the curves are detailed in \Cref{tab:nmin}.

\begin{table*}[!htpb]
\centering
\begin{tabular}{|c|c|c|c|c|c|}
\hline
 \multicolumn{6}{|c|}{$n_{\textrm{min}}$ } \\
 \hline
$F_{i}$ & 0.83 & 0.85 & 0.90 & 0.95 & 0.99 \\
\hline
PGRAND  & 251 & 60   & 16   & 10   & 8    \\
\hline
Hashing $\delta_{\textrm{optimal}}$ & 2326 & 637 & 153  &  71 & 45  \\
\hline
Hashing $\delta_{\textrm{reference}}$ & 8116 & 2027 & 412  & 164  & 82  \\
\hline
\end{tabular}
\caption{Minimum number of pairs required to achieve purification for an ensemble with fidelity $F_{i}$. While choosing an optimal value for $\delta$ greatly reduces the number of required pairs, PGRAND can reduce this number up to 10 times.}
\label{tab:nmin}
\end{table*}

There is a clear exponential relationship between $F_{i}$ and $n_\textrm{min}$, and as expected, the higher the initial fidelity the fewer initial pairs are required to achieve purification. For example, while for achieving purification with $F_{i} = 0.850$ it is required that $n_{\textrm{min}} = 60$, for $F_{i} = 0.818$ we have that $n_{\textrm{min}} = 1947$. Nevertheless, employing as few as 10 qubits enables the purification of states with $F_{i} = 0.95$, while increasing the qubit count to 16 lowers this threshold to $F_{i} = 0.90$. This is in striking contrast with the hashing protocol, which requires at least an ensemble with a size in the order of the hundred qubits for most noise regimes.

For a practical scenario, it is crucial to assess the performance of both protocols under equivalent computational constraints. This entails examining how they fare when correcting a comparable number of errors. However, while PGRAND's performance improves with the number of errors it aims to correct (set by the parameter $t$), the hashing protocol doesn't necessarily follow the same trend. Increasing the number of errors under consideration (represented by a lower value of $\delta$) may result in a decline of its performance, as detailed in \Cref{app:ts}, making direct comparisons challenging. 

\begin{figure}[t]
    \includegraphics[width=\columnwidth, left]{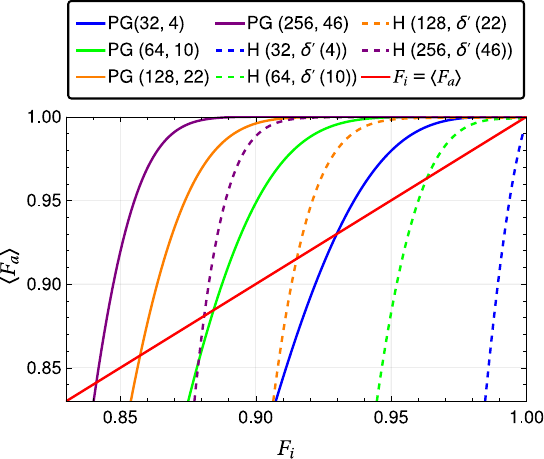}
    \caption{Fidelities $\langle F_{a} \rangle$ by the hashing protocol with $n$ pairs and parameter $\delta$ (denoted by H($n$, $\delta$)) and PGRAND with $n$ pairs and maximum weight $t$ (denoted by PG$(n,t)$)}  as a function of the fidelity of Werner pairs $F_{i}$, when constrained to correct the same number of errors. The solid red line indicates the range where purification is achievable.   
    \label{fig:fidelity_achieved_hashing}
\end{figure}

Nonetheless, we define $\delta'(t) = \frac{1}{n}\log_{2}(N_{\leq t})-S(F_{i})$ and compare the fidelity achieved by the hashing protocol using $\delta' (t)$ with the fidelity attained by PGRAND when limiting the error correction to errors with weight up to $t$. The choice of a suitable value for $t$  is made to ensure a fair and balanced comparison with PGRAND under similar error correction conditions. For both protocols, we considered the scenario where only one purified pair is obtained ($k = 1$). Therefore the results depicted in \Cref{fig:fidelity_achieved_hashing} serve only to illustrate how both protocols perform under the same computational constraints.

We observe that for a small number of pairs, PGRAND exhibits superior performance compared to the hashing protocol. As the number of pairs in the initial ensemble increases, the performance differences between the two approaches tend to diminish. However, it is crucial to consider the computational cost associated with larger initial ensembles. The hashing protocol becomes impractical due to this increased computational burden, while PGRAND maintains the advantage of achieving purification with a smaller number of pairs and manageable computational resources.

PGRAND appears to offer a distinct improvement over the hashing protocol. However, its implementation demands more gates and qubits compared to the hashing protocol; while both require $\mathcal{O}(n^{2})$ gates, the hashing protocol does not use ancilla qubits. Consequently, the advantage of using PGRAND might be diminished if one considers the noise introduced by each gate.

\subsection{Recurrence Protocols}
\label{sec:comparison_recurrence}

% Comparing deterministic protocols such as ours with recurrence protocols has been an underexplored subject in the literature due to the fact that they were designed to be applied in different regimes and constructing good metrics to evaluate them both is still an open problem.
The comparison between deterministic protocols, such as ours, and recurrence protocols remains relatively unexplored in the literature. This is primarily due to their design for application in different regimes, rendering the construction of effective metrics for their evaluation an ongoing challenge in the field.
Some insights about this class of protocols are provided in \Cref{app:ReviewRecurrence}.

Due to their inherent probabilistic nature, our yield definition is no longer appropriate since it does not take into account the possibility of failure of these protocols, in the sense that it does not provides a good insight about the final cost of purification in terms of used Bell pairs, (which is given by $n/P_{\textrm{suc}}$, with $P_{\textrm{suc}}$ being the probability of success of the protocol. For example, according to our definition the Oxford protocol would have a constant yield of $\frac{1}{2}$ for any given initial fidelity. Furthermore for recurrence protocols the expected yield and the achieved fidelity is automatically determined by the protocol, the initial state, and its fidelity. However, in hashing protocols one can manipulate the yield in order to obtain different fidelities. Thus, in order to be able to compare both classes of these protocols, we introduce a new type of yield, which we name the effective yield $Y_{E}$, and define it as 

\begin{equation}
\label{eq:effective_yield}
    Y_{E} =\frac{k}{n/P_{\textrm{suc}}}(F_{a} - F_{i}) = 
    P_{\textrm{suc}} Y \Delta F \in [0,1].
\end{equation}

The effective yield offers a comprehensive assessment metric for purification protocols by integrating both yield and fidelity increase aspects. While lacking a direct physical interpretation, $Y_E$ effectively quantifies the efficiency of protocols in terms of yield and fidelity enhancement, constrained to non-negative values under the assumption that purification protocols cannot diminish fidelity. However, $Y_{E}$ prioritizes the purification cost of a protocol over fidelity increase, particularly accentuated in high initial fidelity regimes where minimizing purification cost is critical. As such, $Y_{E}$ serves as a valuable tool for identifying protocols that strike an effective balance between fidelity enhancement and purification cost reduction, addressing a significant gap in the evaluation of entanglement purification protocols.

% -This metric takes into account both the yield and the fidelity increase.
% - It possess no physical intrepertation 
% - Can only take positive values if we assume that no purification protocol can decrease the fidelity of the essemble
% - The purification cost of a protocol as a major weight than the increase in fidelity, mainly for high values of initial fidelit ( low entropy regime)

\begin{figure}[h]
    \centering
    \includegraphics[width=\columnwidth]{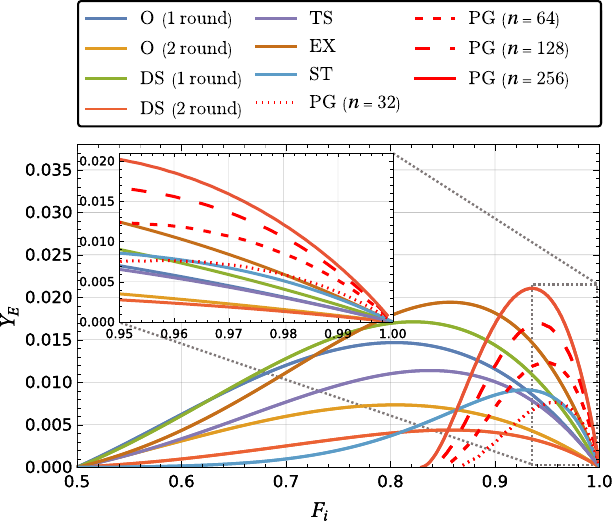}
    \caption{Effective yield $Y_{E}$ of different protocols (Oxford (O), Double Selection (DS), Triple Selection (TS), EXPEDIENT (EX), STRINGENT (ST) and PGRAND (PG)) as a function of the initial fidelity $F_{i}$ of a Werner State, applied in a noiseless scenario.}    
    \label{fig:ComparisonRecurrence2}
\end{figure}

The results for the various protocols regarding the effective yield, when considering an initial ensemble composed by Werner states with fidelity $F_{i}$ can be seen in \Cref{fig:ComparisonRecurrence2}. While for computing $Y_{E}$ for the case of PGRAN the probability of error to has been used to lower bound the fidelity achieved, for the remaining protocols this quantity was directly computed without approximations, meaning that the comparison is in favor of the recurrence protocols.

%By inspecting the resullts, we notice that increasing the number of rounds does not appear to increase the effective yield. This naturally comes as a consequence of the fact that, as it is defined, the effective yield takes more into account the purification cost rather than the increse in fidelity.
% Expedient and Triple Selection seem to be some of the best protocols regarding this metric in a high entropy regime. This is mainly due to the fact that in the high entropy regime, a larger fidelity increase can be obtained.
% PGRAND dominates in the low entropy regime, and the advantage increases with the size of the intial emsemble.
% In this regime, the main contribution to this the effective yield stems from the reduced purification cost, where PGRAND notabl outperforms the other protocols, rather than the increase of fidelity.

Upon examining the findings depicted in the figure, it becomes apparent that the increment in the number of rounds does not necessarily correspond to an elevation in the effective yield. This observation is intricately linked to the definition of the effective yield, which places a greater emphasis on the purification cost rather than the fidelity enhancement.

In high entropy regimes, EXPEDIENT and Triple Selection protocols (see \cref{app:ReviewRecurrence}) emerge as optimal choices in terms of this metric. This superiority can be attributed to the potential for achieving a significant fidelity increase within a high entropy regime, while maintaining reasonable purification costs.

Conversely, in low entropy settings, PGRAND exhibits dominance, with its advantage escalating proportionally with the initial ensemble's size. Within this regime, the primary determinant of the effective yield lies in the diminished purification cost, where PGRAND notably outperforms other protocols, rather than in the fidelity improvement.

% Hence, we introduce the Expected Yield as the reason between the expected number of required initial Bell pairs and the output Bell Pairs for one success atempt of the protocol:

% \begin{equation}
%     Y_{E} = \frac{k}{n/ P_{\textrm{suc}}} = P_{\textrm{suc}} Y
% \end{equation}

% From the above definition, it is clear that $Y \leq Y_{E}$, with the equality being verified only by deterministic protocols of probabilistic protocols applied to states with fidelity equal to one.

\section{Measurement-based version of PGRAND}
\label{sec:measurement}

During the analysis of the protocol we assumed that the operations were noiseless. Nonetheless, this is an unrealistic approach, and even a slight amount of noise can compromise the entire protocol, since the entropy of the ensemble is amplified by each noisy operation. Considering that $\mathcal{O}(n^{2})$ two-qubit operations are required, the increase in entropy caused by noisy operations can easily exceed the information obtained from the measurement. This holds even for minor imperfections, posing a threat to the entire protocol.

Although Clifford gates can be implemented in a fault tolerant manner, this problem can be circumvented by using a measurement-based approach. The main concept revolves around the preparation of a resource state using measurement-based methods, which incorporate both the encoding, stabilizer measurement, and decoding of the protocol. 
Afterward, the Bell pairs are coupled to the resource state using Bell measurements \footnote{Executing the purification protocol in this manner renders it non-deterministic due inherent probabilistic nature of Bell measurements.}, in a manner similar to what is carried out in Refs.~\cite{zwerger2012measurementbased, zwerger2013universal, zwerger2014robustness, zwerger2016measurementbased, miguel-ramiro2018efficient, yan2021feasible, yan2021measurementbased}.  

For an entanglement purification protocol that uses only Clifford gates to distill $k$ pairs from an initial ensemble of $n$ pairs, a minimal resource state can be obtained with $2(n+k)$ qubits. Although the measurement-based approach requires more qubits, ancilla qubits are not needed, as explained in \cite{raussendorf2009measurementbased,nielsen2005faulttolerant}. Additionally, any Clifford circuit can be implemented using only one \textit{measurement layer}, that is, with all measurements being parallelizable. However, it is essential to note that, despite the resulting state is no longer a cluster state (and therefore is no longer universal) it is, up to local unitaries, equivalent to a graph state \cite{raussendorf2009measurementbased, hein2004multiparty}.

The resource states can be obtained from a cluster state and then purified and encoded. Quantum communication via measurement-based entanglement purification has been shown to outperform direct transmission, gate-based error correction, and measurement-based schemes generating resource states directly \cite{wallnofer2017measurementbased}.
 
The fact that the resource state can be prepared in advance is especially relevant because qubits may experience dephasing during the processing time of the protocol. However, with the resource state prepared beforehand, the processing time is uniquely limited to the time required to perform the Bell measurements and classically communicate the results.

 \begin{figure}[t]
    \centering
    \includegraphics[width=0.49\textwidth]{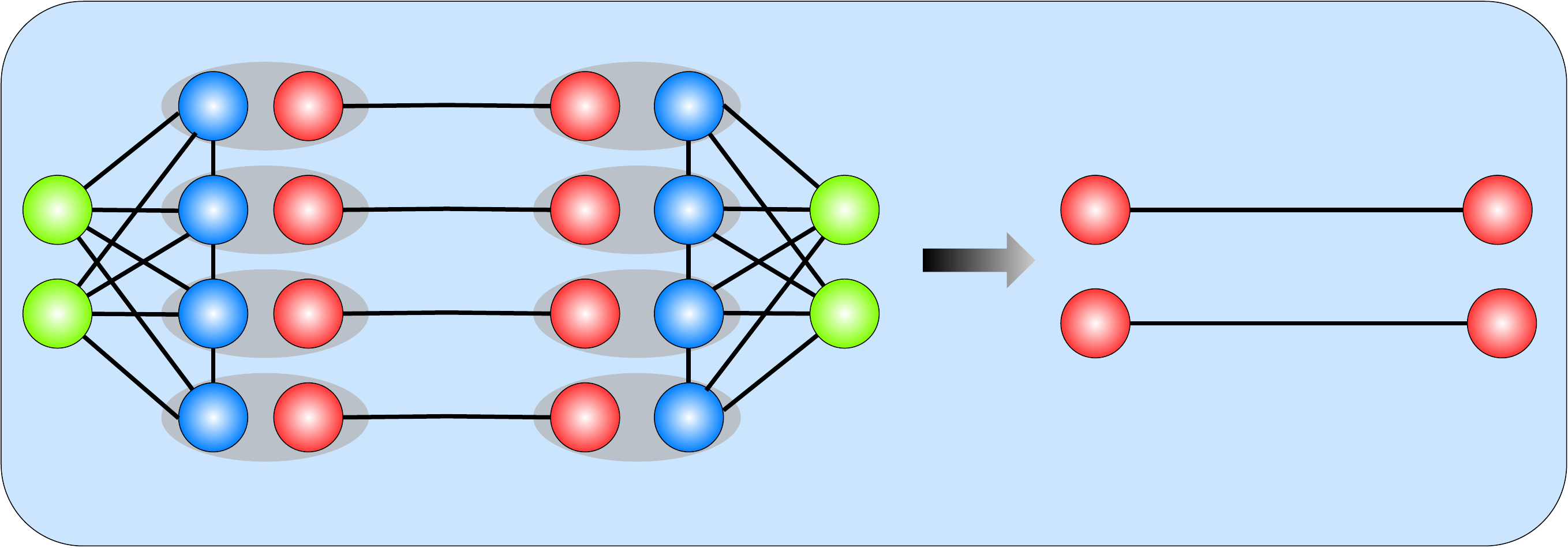}
    \caption{Illustration of the implementation of the protocol in a measurement-based manner. The Bell pairs (red vertices) are coupled to the input qubits of the resource state (blue vertices) via Bell Measurements. The output qubits of the resource state (green vertices) are kept and the remaining ones are discarded. After communicating classically the measurement results and applying the corrections, purified Bell pairs are obtained. }    
    \label{fig:MBHashing}
\end{figure}

To explore how imperfections in quantum operations and noisy Bell measurements affect a measurement-based entanglement purification protocol, specifically, the impact of imperfect resource states,  we consider LDN on each particle of the resource states. For estimating the error thresholds in a noisy implementation for this protocol, we remark that it has been proven that it is possible to formally transfer the noise affecting the input qubits of the resource state to the Bell pairs targeted for purification \cite{zwerger2013universal}.  Consider that the Bell pairs and the resource states are affected by LDN with parameters $p$ and $q$ respectively.

\begin{figure}[t]
    \centering
    \includegraphics[width=\columnwidth]{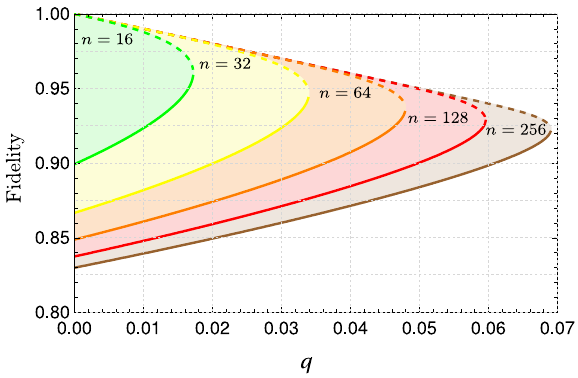}
    \caption[Maximum achievable fidelity (dashed lines) and minimum required fidelity (solid lines) of PGRAND as a function of the noise parameter $q$ of the resource states]{Maximum achievable fidelity (dashed lines) and minimum required fidelity (solid lines) of PGRAND as a function of the noise depolarization probability $q$ of the resource states, for $n$ qubits. The colored space between the solid and the dashed lines of the same color represents the purification range of the protocol, meaning that in these areas purification is achievable.}    
    \label{fig:error_thresholds}
\end{figure}

It is easily verifiable through simple computations that this is equivalent to considering that one starts with Werner states with fidelity
\begin{equation}
    F_{i} = 1 - p - q + \frac{4}{3}pq.
\end{equation}
Similarly, the noise present in the output qubits can also be transferred to the final state, given that one can always assume that the protocol outputs Werner states with fidelity $\langle F_{a}\rangle$. Thus, after moving the noise from the output qubits of the resource state to the output Bell pairs, one ends up again with a Werner state with fidelity
 \begin{equation}
     \langle F_{a} \rangle = 1- p_{e}- q + \frac{4}{3}p_{e}q.
 \end{equation}
 The advantage of this decomposition is that, once the noise has been transferred, we are left with a perfect purification process to which our previously presented analytical results still apply. 
With that in mind, we can now deduce a threshold for the tolerable error value of the resource states. By recognizing that it is impossible to purify Bell pairs with a fidelity lower than that of the resource state, we obtain the condition $p \geq q$. This condition arises from the requirement that the purification process must be achievable, which imposes $F_{a} \geq F_{i}$. Moreover, the protocol can only operate on a depolarized state with fidelity $F_{i} \geq F_{\textrm{min}} = 0.8107$. Hence the following condition is obtained 
\begin{equation}
    1-2q+\frac{4}{3}q^{2}\leq F_{\textrm{min}},
\end{equation}
implying the threshold to be $q_{\textrm{min}} = 0.0859$.  
Figure \ref{fig:error_thresholds} presents a numerical analysis of the purification ranges corresponding to various numbers of initial pairs, while \Cref{tab:error_thresholds} displays the error thresholds for the same numbers of initial pairs. These thresholds for the minimum required fidelity $F_{\textrm{min}}$ and the maximum achievable fidelity $F_{\textrm{max}}$ were obtained by inspecting the $n$-to-one routine. These values can be seen in \Cref{tab:min_fidelity} and
\Cref{tab:max_fidelity}, respectively. 
It is important to highlight that these error thresholds align with a reasonable and conservative error model consistent with measurement-based quantum information processing principles. However, it is not straightforward to compare these thresholds directly with those derived from gate-based error models, such as the ones used in \cite{deutsch1996quantum,krastanov2019optimized}. Establishing such a connection presents an open challenge. Furthermore, extending the analysis to include correlated errors would be valuable. This extension could involve exploring how errors are localized and justifying a local error model through the application of multipartite entanglement purification for resource states \cite{zwerger2012measurementbased, aschauer2005multiparticle, dur2003multiparticle}. 

\begin{figure}[!t]
    \centering
    \includegraphics[width=\columnwidth]{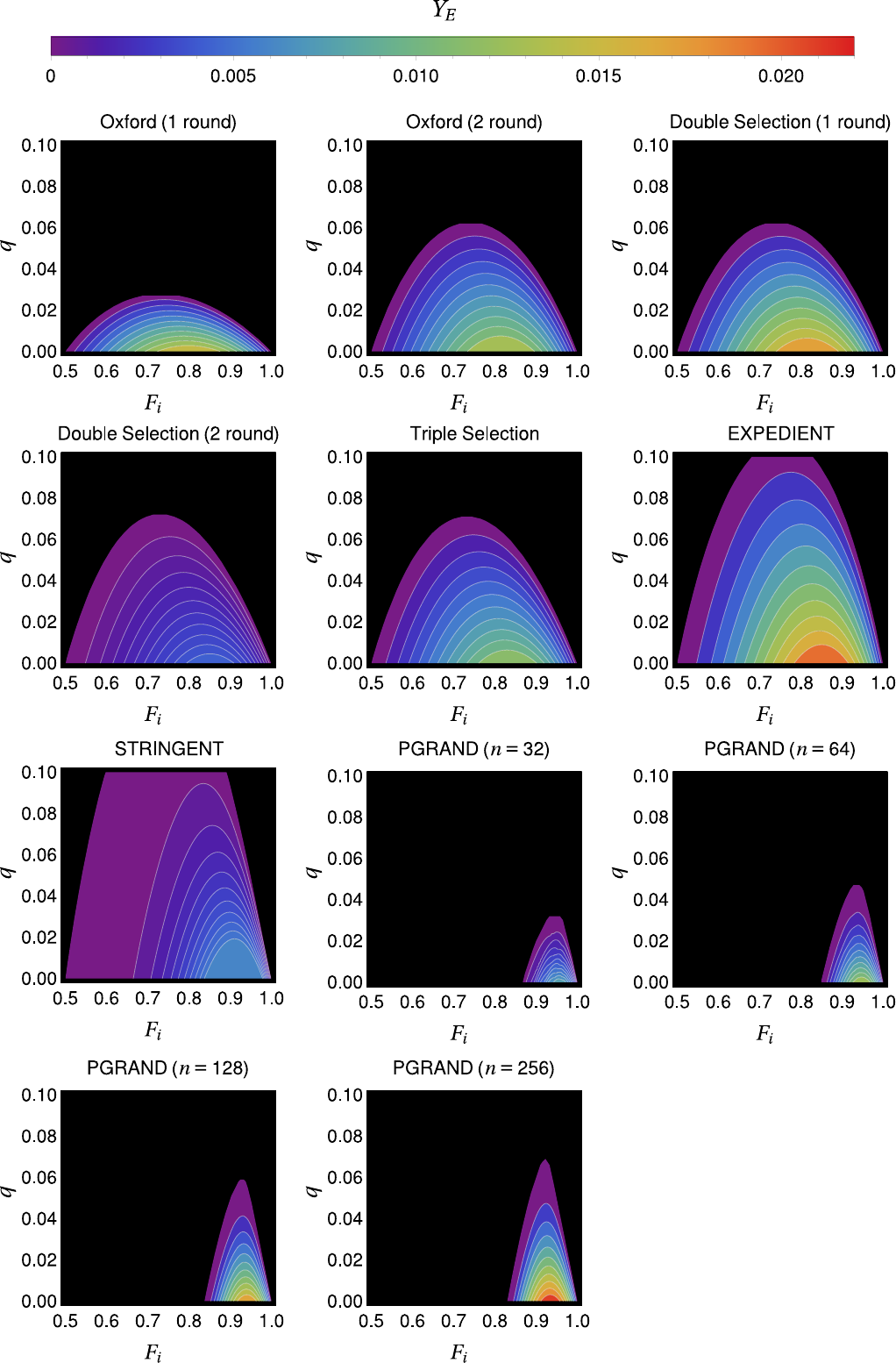}
    \caption{Contour plots of the effective yield $Y_{E}$ for different protocols as a function of the initial fidelity $F_{i}$ of a Werner State and the parameter $q$ of LDN in the resource states. The black area denotes the zone where purification is not achieved.}    
    \label{fig:ComparisonRecurrence3}
\end{figure}

\begin{table}[!t]
\centering
\begin{tabular}{|c|c|c|c|c|c|}
\hline
 \multicolumn{6}{|c|}{Error Threshold } \\
 \hline
$n$ & 16   & 32   & 64   & 128  & 256  \\
\hline
  $q$  & 1.72\% & 3.39\% & 4.79\% & 5.96\% & 6.89\% \\
\hline
\end{tabular}
\caption{Resource states error thresholds as a function of the number of required Bell pairs, expressed by the LDN parameter $p$ per qubit. All of these results are referent to the $n$-to-one routine.}
\label{tab:error_thresholds}
\end{table}

In order to establish a comparison, one can also apply the explained formalism to create measurement-based versions of the purification protocols presented in \Cref{sec:comparison_recurrence}. The noise of the resource states can be modelled in the same way. Just like in the previous scenario, we used the effective yield as a measure of performance of the protocols.

The results are shown in the form of contour plots in \Cref{fig:ComparisonRecurrence3}. For the cases where 2 rounds of a recurrence protocol were applied, we considered resource states of reduced size that apply the purification procedures within a single step  \cite{zwerger2016measurementbased}. Notice that this results in the increase of the contour area, since purification can be achieved more easily given that more rounds allow greater fidelity increases. This comes with the trade off of diminishing the efficient yield due to lower success probabilities, which ultimately translates into higher purification costs.

The results also indicate recurrence protocols to be much more resilient to the effect of noise, as it has already been pointed out in \cite{zwerger2013universal}. This is attributed to those protocols' larger tolerance to initial noise, given that they are able to theoretically tolerate any $F_{\textrm{min}} \geq 0.5 $, while PGRAND and the hashing protocol require $F_{\textrm{min}} \geq 0.8107 $. EXPEDIENT seems to be the best performing protocol according to the effective yield for the widest ranges of $F_{i}$ and $q$. Nonetheless, an advantage of using PGRAND can be seen again in the low entropy scenario ( $q < 0.001$ and $F_{i} > 0.90$)), and the advantage is more prominent as the size of the initial ensemble grows.

%\newpage
\section{Conclusions}
\label{sec:conclusion}
Quantum entanglement purification methods are vital in quantum information processing. They help mitigate noise and enable the recovery of highly entangled states, making them essential tools for various quantum information tasks.
In this paper, we present a purification protocol that represents a significant improvement over Bennett's hashing protocol. We provide a detailed explanation of the protocol's construction concepts and the underlying principles, along with a comprehensive analysis of its performance across different ensemble sizes, noise regimes, and computational constraints.
Our findings demonstrate that our protocol can achieve a fidelity of 90\% when purifying states, even with a relatively small ensemble size of just 16 Bell pairs. While hashing protocols offer distinct advantages over other types of protocols due to their non-probabilistic nature (unlike recurrence protocols) and high performance as the initial ensemble size increases, they are constrained by the computational resources they require \cite{chau2011practical,zwerger2018longrange, zwerger2014robustness}. 
The requirement of such a small initial ensemble in our protocol not only reduces computational costs significantly but also paves the way for a more realistic implementation. This is especially relevant in practical scenarios where there is a preference for employing fewer qubits. Moreover, if one manages to obtain a larger ensemble of Bell pairs with sufficiently low levels of noise, high efficiencies can be obtained while maintaining a feasible computational cost. Thus, PGRAND purports to be specially relevant in low-entropy noise regimes.
The assumption of using noiseless gates can be overcome by using a measurement-based implementation of these protocols. Furthermore, it is known that classical RLCs can reach Shannon's channel capacity and are easy to generate. Low-density parity-check (LDPC) codes are also capacity-achieving when decoded with the belief propagation algorithm, but their construction imposes some constraints and not all $(n,k)$ pairs may be feasible. A fundamental characteristic of LDPC codes is the sparsity of the parity-check matrix. In a quantum setup using noise-guessing framework with a syndrome-based membership test, this sparsity induces a syndrome-checking procedure that involves many fewer gates than the one using QRLCs. For this reason, for fault-tolerant syndrome computation, the use of LDPC codes may be preferable when using the proposed PGRAND, given their more limited exposure to gate errors. The general fault-tolerance of a gate implementation of QGRAND/PGRAND is subject of ongoing work \textcolor{red}{\cite{cruz2024fault}}.
The extension of our protocol to multipartite entanglement is also an interesting question to study. Given that the Bell matrix identity can generalized to GHZ states \cite{rengaswamy2022distilling}, a version of the protocol to these states should be conceivable. Furthermore,
the performance of our protocol could be improved by considering its generalization to multi-level systems (qudits), since notable improvements were found for a bipartite qudit implementation of the hashing protocol \cite{miguel-ramiro2018efficient}. It would also be worthy to study in a more detailed scenario the feasibility of the measurement-based of this protocol, and how the cluster state size/complexity would influence the error of the resource states in a realistic implementation. 

\section*{Acknowledgments}
Francisco Monteiro and Bruno Coutinho are grateful to Prof.  Wolfgang D\"ur (University of Innsbruck)  for insightful discussions about hashing-based purification.
This work was supported by FCT - Fundação para a Ciência e Tecnologia, I.P. by project reference UIDB/50008/2020, and DOI identifier \url{https://doi.org/10.54499/UIDB/50008/2020}, and by project reference QuNetMed 2022.05558.PTDC, and DOI identifier \url{https://doi.org/10.54499/2022.05558.PTDC}.
Diogo Cruz acknowledges the support from FCT through scholarship UI/BD/152301/2021.

%\bibliography{./References.bib}
%\begin{verbatim}
\bibliographystyle{quantum}
\bibliography{References.bib}
%\end{verbatim}

\appendix

\section{Probability of error as a lower bound of the fidelity}
\label{app:bler}

We consider the probability of error $p_{e}$ to be the ratio between the number of successful uses of that protocol and the total number of uses. This can be interpreted as if any uncorrected error pattern maps the initial state to an orthogonal state, that is, $\mathcal{F}( | \psi \rangle, E | \psi \rangle) = 0$, for any uncorrected error $E$. If $\mathcal{E}$ is a set of $N$ possible error patterns, and $\mathcal{C} \subseteq \mathcal{E}$ is the subset of correctable errors, then the average achieved fidelity $\langle F_{a} \rangle$ of the state obtained at the end of the protocol verifies:
\begin{align}
\label{eq:bler}
    \langle F_{a} \rangle & = \frac{1}{N} \sum_{E_{i} \in \mathcal{E}} \mathcal{F}( | \psi \rangle , E_{i} | \psi \rangle) = \\
    &\frac{1}{N} [ \sum_{E_{i} \in \mathcal{C}}
        \mathcal{F}( | \psi \rangle, E_{i} | \psi \rangle) + 
        \sum_{E_{i} \in \mathcal{E} \setminus \mathcal{C}} \mathcal{F}( | \psi \rangle, E_{i} | \psi \rangle) ) ] \\
        & \geq \frac{1}{N} \sum_{E_{i} \in \mathcal{C}}
        \mathcal{F}( | \psi \rangle, E_{i} | \psi \rangle) = 1 - p_{e}
\end{align}

\noindent since it is considered that we are able to correct perfectly every error in $\mathcal{C}$, that is,  $ \forall_{E_{i} \in \mathcal{C}} \; \hspace{2mm} \mathcal{F}( | \psi \rangle, E_{i} | \psi \rangle) = 1 $.  This is equivalent to model the output states of the protocol as Werner states with fidelity  $\langle F_{a} \rangle $.

% $Y = $ yield

% For recurrence protocols the expected yield and the achieved fidelity is automatically determined by the protocol and the initial fidelity of the Bell pairs. However, in hashing protocols one can manipulate the yield in order to obtain different fidelities. Thus, in order to be able to compare both classes of these protocols, we introduce a new metric, which we name the Purification Efficiency Index, and define it as 

% \begin{equation}
%     Y_{E}(F_{a} - F_{i}) = Y_{E}\Delta F \in [0,1]
% \end{equation}

\section{Suboptimality of the hashing protocol}
\label{app:ts}

\begin{figure}[!t]
    \centering
    \includegraphics[width=\columnwidth]{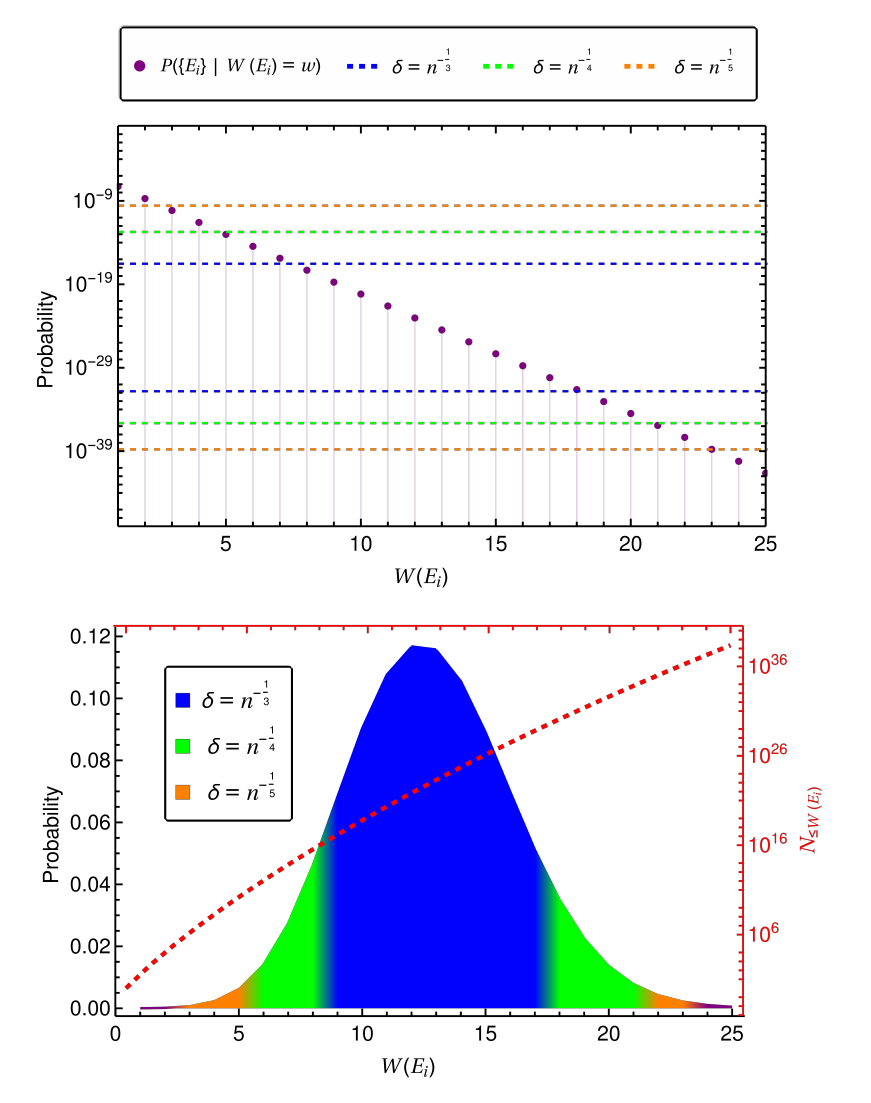}
    \caption{This figure illustrates how different choices of the $\delta$ parameter affect the correction capability of the hashing protocol. In plot a), each dot represents the individual probability of an error pattern based on its weight. Each pair of dashed lines of the same color represents the bounds imposed by the choice of $\delta$ (as defined in \Cref{eq:ts_bound}). In plot b), we observe the sum of the individual probabilities of error patterns with the same weight. The colored area represents the total probabilities of all errors included in the typical set, depending on the value of $\delta$. The red line indicates the number of error patterns at each weight. From this plot, it becomes evident that the correction procedure of the hashing protocol is never optimal, regardless of the choice of $\delta$. In both plots, we consider a scenario with an initial ensemble of 128 pairs subject to 10\% LDN.}    
    \label{fig:typical_set}
\end{figure}

Consider the typical set as defined in \Cref{eq:ts}. Assuming that $W$ corresponds to the density matrix of a Werner state with parameter $F$, the typical set encompasses all error patterns for which their probabilities satisfy

\begin{equation}
\label{eq:ts_bound}
2^{-n(S(F)+\delta)} \leq P({E_{i}}) \leq 2^{-n(S(F)-\delta)}.
\end{equation}
This set contains at most $2^{n(S(W)+ \delta)}$ error patterns whose weight is closer to $n(1-F)$. As defined, the hashing protocol treats all errors within this set as equally important, implying that any error pattern in this set has the same probability of being corrected. However, this correction procedure is suboptimal, regardless of the choice of $\delta$.

On one hand, a high value for $\delta$ will inherently limit the protocol's performance, since even if it manages to correct all errors within the typical set, the total probability of error is upper-bounded by the sum of probabilities of errors that are not contained within the set. On the other hand, choosing a small value for $\delta$ implies the inclusion of many errors with relatively high weight. As the number of error patterns exponentially increases with their weight, it becomes more likely (for a fixed number of syndromes) that the corrected errors will be the ones with greater weight. However, each of these errors contributes only a small amount to the overall error probability of the protocol. In \Cref{fig:typical_set}, a practical scenario illustrates how the choice of $\delta$ interferes with the hashing protocol's performance.
For illustration purposes, we have selected an ensemble containing 128 qubits with 10\% LDN, but the same holds true for other choices. Even with an optimal choice of $\delta$ and a sufficiently large $n$, the correction procedure is never optimal. If we consider the computational cost of mapping each syndrome to an error pattern, it becomes evident how the hashing protocol will often fail to achieve purification for most practical cases.

\section{Review of Recurrence Protocols}
\label{app:ReviewRecurrence}

Bipartite recurrence protocols are characterized by their operation within each purification step, which entails processing a fixed number $n$ of copies of a mixed state. After local manipulation, one of these copies is preserved while the remaining $n-1$ undergo measurement. Depending on the measurements outcomes, the remaining copy is either retained or discarded. If the purification step is successful, one obtains a Bell pair with increased fidelity. This iterative procedure continues, using states resulting from successful purification rounds  as input for subsequent iterations. Generally, these protocols converge to a maximally entangled state \cite{dur2007entanglement}.

\begin{figure}[t!]
    \centering
        \includegraphics[width=\columnwidth]{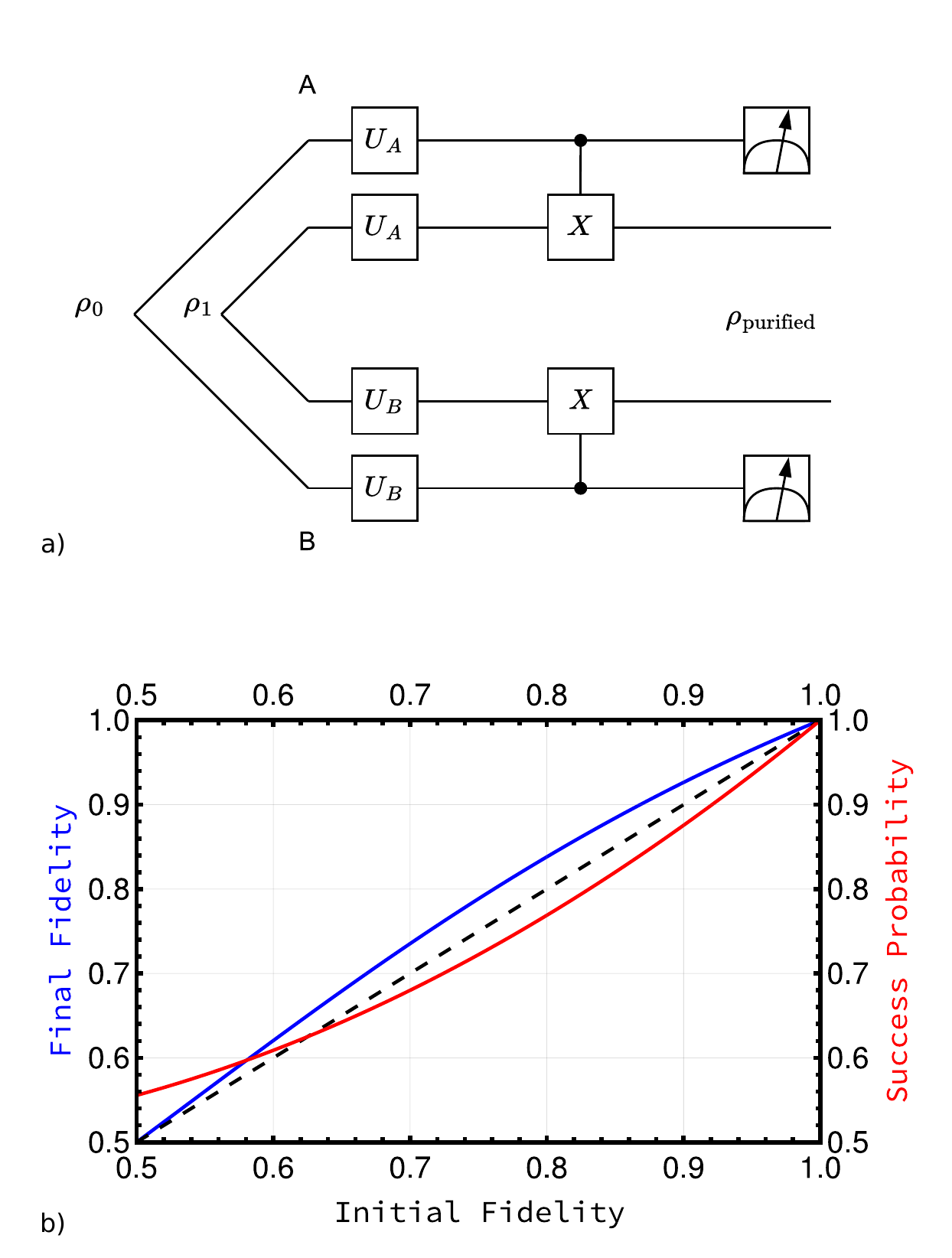}
    \caption{The Oxford Protocol. On a), the circuit required for the Oxford protocol. As long as the measurement results agree, a purified Bell pair is obtained.} On b), the plot of the fidelity obtained and success probability of the Oxford protocol when applied to two Werner states in relation to its initial fidelity.
    \label{fig:oxford_protocol}
\end{figure}

Within this class of protocols, one of the simplest (yet efficient) protocols is the Oxford protocol \cite{deutsch1996quantum}, a recurrence protocol that is conceptually similar the IBM protocol \cite{bennett1996mixedstate, bennett1996purification}. Consider that two noisy Bell pairs are shared between Alice and Bob. They start by applying  the following unitary operations to both of their qubits
\begin{align}
\label{eq:Oxford}
    & |0 \rangle \xrightarrow{U_{A}} \frac{1}{\sqrt{2}} (|0\rangle - i |1\rangle ), \quad |1 \rangle \xrightarrow{U_{A}} \frac{1}{\sqrt{2}} (|1\rangle - i |0\rangle ), \\
    & |0 \rangle \xrightarrow{U_{B}} \frac{1}{\sqrt{2}} (|0\rangle + i |1\rangle ), \quad |1 \rangle \xrightarrow{U_{B}} \frac{1}{\sqrt{2}} (|1\rangle + i |0\rangle ).
\end{align}
after which they both apply a CNOT gate on their qubits. Each measures the target qubit of the CNOT gate, and the measurement results are shared. If the measurement results agree, the unmeasured pair is kept, otherwise it is discarded. The circuit required by this protocol is depicted in \Cref{fig:oxford_protocol} a).

This procedure can be repeatedly applied to the purified pairs. However, at the beginning of each round, the pairs are not required to be Werner states. Instead, one considers states of the general form
\begin{align}
\label{eq:djmps_form}
\nonumber
    \rho = &A |\phi^{+}\rangle\langle\phi^{+}| +
    B |\psi^{-}\rangle\langle\psi^{-}|+
    C |\psi^{+}\rangle\langle\psi^{+}|+\\
    &D |\phi^{-}\rangle\langle\phi^{-}| 
    + \textrm{off-diagonal terms}.
\end{align}
After $r+1$ successful rounds of purification, the following state is obtained
\begin{align}
\label{eq:djmps_purified_form}
\nonumber
    &A_{r+1}=\frac{A_{r}^2+B_{r}^2}{N_{r}}, \quad  B_{r+1}=\frac{2C_{r}D_{r}}{N_{r}}, \\
    &C_{r+1}=\frac{C_{r}^2+D_{r}^2}{N_{r}}, \quad D_{r+1}=\frac{2A_{r}B_{r}}{N_{r}}.
\end{align}
where $N_{r} = (A_{r}+B_{r})^{2} + (C_{r}+D_{r})^{2}$ is equal to the probability of a successful purification round (i.e, the probability that Alice and Bob measurements agree).
As opposed to the IBM protocol, this protocol has the advantage of not requiring a depolarization procedure between subsequent rounds of purification, leading to higher efficiencies \cite{deutsch1996quantum,aschauer2005quantum}. Furthermore,  it has been shown that this procedure converges to the fixpoint defined by ($A_{\infty} = 1, B_{\infty}= C_{\infty}= D_{\infty} = 0 $) if and only if the initial fidelity (for Werner states, this corresponds to the parameter $A$) is greater than $\frac{1}{2}$ \cite{macchiavello1998analytical}.

This protocol has been generalized by \cite{fujii_entanglement_2009} to be applied to $n \geq 2$ Bell pairs and obtain a purified one. The Double Selection and Triple Selection protocols are simply the instances with $n = 3$ and $n = 4$. By increasing the number of auxiliary pairs, the amount of errors that go undetected by the protocols is reduced, resulting in higher fidelities of the resulting state. However, this naturally comes with the downfall of reducing the probability of success of each purification round.

\begin{figure}[t!]
    \centering
    \includegraphics[width=\columnwidth]{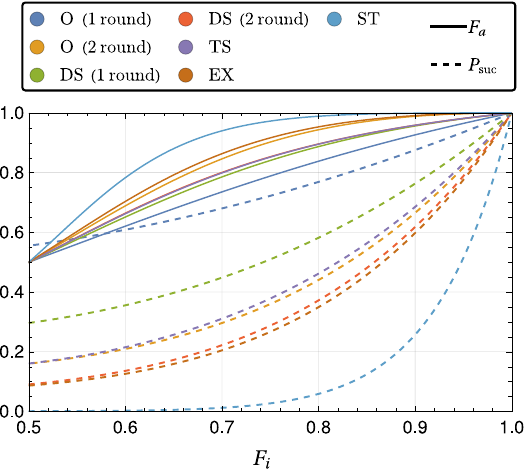}
    \caption{Probability of success $P_{\textrm{suc}}$ and fidelity achieved $F_{a}$ for the considered recurrence protocols (Oxford (O), Double Selection (DS), Triple Selection (TS), EXPEDIENT (EX) and STRINGENT (ST)) when applied to Werner states with an initial fidelity $F_{i}$. A tradeoff can be seen between these two quantities. For a recursive application of a given protocol, increasing the number of purification rounds increases the achieved fidelity at the cost of reducing the probability of success.} 
    \label{fig:recurrence_protocols}
\end{figure}

The other two protocols present in this paper, EXPEDIENT and STRINGENT, are 3-to-1 protocols that were developed within the context of topological quantum computing \cite{nickerson_topological_2013}. The circuit required to implement STRINGENT can be obtained from EXPEDIENT by adding some gates and single-qubit measurements. While STRINGENT is characterized for achieving high fidelities at the cost of low success probabilities, as it can be seen in \Cref{fig:recurrence_protocols}, both are some of the best performing purification protocols \cite{krastanov2019optimized}. All of these protocols can be applied recursively, but increasing the number of purification rounds does not necessarily increases the overall protocol performance, since their success probability decreases exponentially with the number of rounds.

\section{Additional numerical results}
\label{app:additional}

\begin{table*}[t!]
\centering
\begin{tabular}{|c|cccccc|}
\hline
 \multicolumn{7}{|c|}{$F_{\textrm{min}}$} \\
 \hline
                    routine& 1\%   & 2\%   & 3\%   & 4\%   & 5\%   & 6\%   \\
\hline
$16$-to-$1$  & .9238 & n/a   & n/a   & n/a   & n/a   & n/a   \\
$32$-to-$1$  & .8818 & .8996 & .9236 & n/a   & n/a   & n/a   \\
$64$-to-$1$  & .8608 & .8749 & .8890 & .9072 & n/a   & n/a   \\
$128$-to-$1$ & .8479 & .8549 & .8711 & .8845 & .9002 & n/a   \\
$256$-to-$1$ & .8394 & .8494 & .8601 & .8713 & .8838 & .8984 \\
\hline
\end{tabular}
\caption{Minimum fidelity required to achieve purification varies according to different degrees of error in the resource states and routines. Routines that use more pairs would in principle imply utilizing resource states with a higher degree of error. However, it is possible to observe that using more pairs can lead to higher purification thresholds.}
\label{tab:min_fidelity}
\end{table*} 
\begin{table*}[t!]
\centering
\begin{tabular}{|c|cccccc|}
\hline
 \multicolumn{7}{|c|}{$F_{\textrm{max}}$} \\
 \hline
                  routine& 1\%   & 2\%   & 3\%   & 4\%   & 5\%   & 6\%   \\
\hline
$16$-to-$1$  & .9878 & n/a   & n/a   & n/a   & n/a   & n/a   \\
$32$-to-$1$  & .9899 & .9795 & .9650 & n/a   & n/a   & n/a   \\
$64$-to-$1$  & .9899 & .9799 & .9699 & .9587 & n/a   & n/a   \\
$128$-to-$1$ & .9899 & .9799 & .9699 & .9599 & .9496 & n/a   \\
$256$-to-$1$ & .9899 & .9799 & .9699 & .9599 & .9599 & .9398 \\
\hline
\end{tabular}
\caption{Maximum achievable fidelity, according to different degrees of error of the resources states and sizes of the initial assemble. The achieved fidelity is primarily limited by the error of the resources states. }
\vspace{1.29cm}
\label{tab:max_fidelity}
\end{table*} 

\begin{figure*}[!!bthp]
    \centering
    \includegraphics[width=\textwidth]{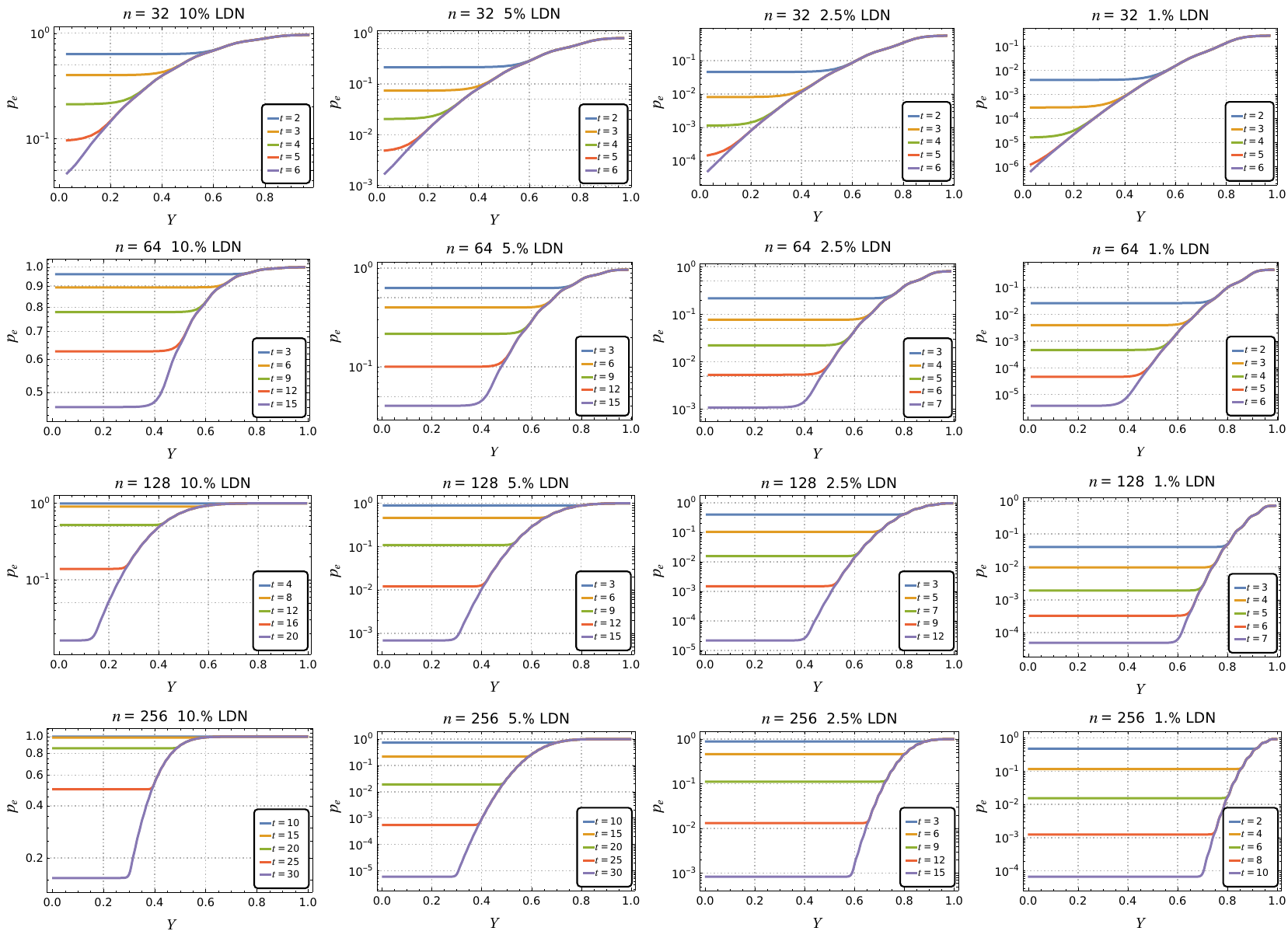}
    \caption{Plot of the error probability $p_{e}$ as a function of the yield for a number $n$ of pairs, where $n = 32$ (first line), $n = 64$ (second line), $n = 128$ (third line), and $n = 256$ (fourth line). Local depolarizing noise was assumed to be 10\% (first column), 5\% (second column), 2.5\% (third column), and 1\% (fourth column). In these plots, it is possible to see how the computational cost, expressed by the parameter $t$, limits the protocol's purification capabilities. The horizontal line on each curve indicates the existence of a saturation point, where most of the considered error patterns were assigned to a syndrome, and no advantage is obtained by sacrificing more pairs (i.e., using lower yields). The computational cost becomes a limiting factor when aiming for purification in scenarios with high amount of noise and/or a high number of pairs. However, for low-entropy regimes ($\leq 2.5\%$ LDN), this cost is still bearable even when considering an ensemble with 256 pairs. From these plots, it is also clear the increase in the protocol's efficiency, expressed by the yield, with the size of the ensemble.}
    \label{fig:bler_theory}
\end{figure*}

In this section, we present complementary numerical results. In \Cref{fig:bler_theory} is possible to see different results regarding the probability of error of the protocol, for different levels of noise, ensemble sizes and computational constraints. These values were obtained solely using the theoretical expressions \Cref{eq:theoretical_fraction} and \Cref{eq:theoretical_bler}.

In \Cref{tab:min_fidelity} and \Cref{tab:max_fidelity} is possible to see the minimum fidelity required for purification, and the maximum achievable fidelity, respectively,  for different degrees of error in the resource states and the sizes of the initial assemblies for the measurement-based version of the protocol.

\end{document}